\documentclass[aps,twocolumn,nopacs,floats,superscriptaddress]{revtex4-2}
\usepackage{graphicx}
\usepackage{dcolumn}
\usepackage{bm}
\usepackage{amsmath}
\usepackage{amssymb}
\usepackage{color}

\usepackage{xcolor}

\definecolor{cardinal}{rgb}{0.6,0,0}
\definecolor{darkgreen}{rgb}{0,0.4,0}
\definecolor{golden}{rgb}{0.92, 0.7, 0}
\definecolor{midnight}{rgb}{0, 0, 0.5}
\definecolor{darkblue}{rgb}{0, 0, 0.7}

\def\he4{$^4$He}
\def\hel3{$^3$He}
\def\Am3{\AA$^{-3}$}
\def\beq{\begin{equation}}
\def\eeq{\end{equation}}

\newcommand{\cH}{{\mathcal H}}

\newcommand{\be}{\begin{equation}}
\newcommand{\ee}{\end{equation}}
\newcommand{\bea}{\begin{eqnarray}}
\newcommand{\eea}{\end{eqnarray}}
\newcommand{\bse}{\begin{subequations}}
\newcommand{\ese}{\end{subequations}}
\newcommand{\dr}{\mathrm{d}}

\begin{document}

\author{Viktor Berger}
\affiliation{Department of Physics, University of Massachusetts, Amherst, MA 01003, USA}

\author{Nikolay Prokof'ev}
\affiliation{Department of Physics, University of Massachusetts, Amherst, MA 01003, USA}

\author{Boris Svistunov}
\affiliation{Department of Physics, University of Massachusetts, Amherst, MA 01003, USA}

\title{ 
``Depletion" of Superfluid Density:\\
Universal Low-temperature Thermodynamics of Superfluids
}

\begin{abstract}
In a Galilean superfluid, the depletion of superfluid density with rising temperature can be attributed to thermally excited non-interacting phonons. For systems without Galilean symmetry, it has been shown \cite{companion} that ``phonon wind" is no longer responsible for the depletion of superfluid density. In this work, we develop the theory of superfluid density at low temperature ($T$) and provide detailed derivations of all results announced in
\cite{companion}. Using Popov's hydrodynamic action, we show that the theory of low-temperature depletion 
in a $d$-dimensional quantum superfluid maps onto the problem of finite-size ($L$) corrections in a $(d+1)$-dimensional anisotropic (pseudo-)classical-field
system with U(1)-symmetric complex-valued action. In addition to generalizing Landau's (canonical) formula, we develop the grand canonical theory, which in a broader context reveals a universal scaling, $T^{d+1}$ and 1/$L^{d+1}$, for finite-$T$ and finite-$L$ effects of many thermodynamic quantities.
We validate our theory with numeric simulations of interacting lattice bosons and the J-current model.
\end{abstract}

\maketitle

\section{Introduction}
\label{sec:Intro}
The theory of superfluidity by Landau \cite{Landau_1941} was revolutionary for the understanding of not only liquid helium, but of condensed matter physics as a whole. Often dubbed a ``macroscopic quantum phenomenon," superfluidity has long served as a prototype example of how the quantum nature of particles can manifest at large scales. Essentially, all the universal properties exhibited by superfluids are best understood in terms of the field of superfluid phase. Remarkably enough, Landau was able to formulate his theory of superfluidity well before the description in terms of the phase field was discovered \cite{Anderson_1966}. Crucially relying on the assumption of Galilean invariance, Landau derived an explicit formula for the low-temperature depletion of the superfluid density. The theory associated the normal component of the superfluid with the phonon wind, the momentum of which is easily calculated through the Gibbs statistics of non-interacting elementary excitations. The Galilean relationship $\mathbf{j}_m=\mathbf{p}$ between the mass density flux and momentum density then immediately yields Landau's formula.

In a companion letter \cite{companion}, we have presented the generalization of Landau's depletion formula for a $d$-dimensional system at fixed number density $n$, taking the form
\be\label{eqn:delta_ns_final}
\Delta n_s(n,\beta)=-\frac{I_d}{d}\frac{\nu^2(d+1)-d\gamma n_s -(d+2)\sigma/\varkappa}{c(c\beta)^{(d+1)}}\,,
\ee
\be\label{eqn:Id_def}
I_d=\int\frac{\mathrm{d}^dq}{(2\pi)^d}\frac{q}{e^q-1}\, ,
\ee
\be\label{eqn:nu}
\nu=\frac{\dr n_s}{\dr n}\,, \quad \gamma=\frac{1}{2}\frac{\dr^2n_s}{\dr n^2}\,, \quad \sigma = 2\frac{\partial^2 \mathcal{E}(n,k_0^2)}{\partial(k_0^2)^2}\bigg|_{k_0=0}\, .
\ee
Here $n_s$ is the superfluid stiffness related to superfluid density $\rho_s$ through $\rho_s=mn_s$ ($m$ being the particle mass), and $\Delta n_s(n,\beta)=n_s(n,\beta)-n_s(n,\infty)$ is the deviation of the superfluid stiffness from it's $T=0$ value. All parameters appearing in Eq.~(\ref{eqn:delta_ns_final}) are ground-state properties of the system, including the sound velocity $c=\sqrt{n_s/\varkappa}$, compressibility $\varkappa=dn/d\mu$ with $\mu$ being the chemical potential, and energy density $\mathcal{E}(n,k_0^2)$ (expressed as a function of number density $n$ and square of superflow wavevector $k_0$). The formula~(\ref{eqn:delta_ns_final}) reproduces Landau's in the Galilean limit where $\nu=1/m$ and $\gamma=\sigma=0$.

It should come as no surprise that the generalization of Landau's theory explicitly invokes the concept of the superfluid phase field. Our theory utilizes Popov's hydrodynamic action \cite{Popov}, where the ground-state quantities in Eq.~(\ref{eqn:delta_ns_final}) enter as expansion parameters in a perturbative expansion of the energy density $\mathcal{E}(n,k_0^2)$. At the harmonic level, calculating finite-$T$ corrections to $\Delta n_s$ is analogous to computing finite-size effects in the compactified imaginary-time direction. Thus, it is instructive to first study finite-size effects in a \textit{classical-}field system with broken $U(1)$ symmetry.

The paper is organized as follows.
In Sec.~\ref{sec:Class}, we consider mapping between the low-temperature
$d$-dimensional quantum superfluid and $(d+1)$-dimensional classical-field
system and how finite-temperature effects in the former relate to finite-size effects in the latter. This allows us to introduce and study 
$n_s$ depletion effects, both theoretically and numerically, in the simplified space-time symmetric case. 
Quantum case is discussed in Sec.~\ref{sec:Quant}. Based on effective hydrodynamic Hamiltonian, we express $\Delta n_s$ in terms of certain correlation functions.  Calculating then the finite-$T$ and finite-$L$ contributions to these correlation functions---with bilinearized hydrodynamic action, we produce the desired results, including 
Eqs.~(\ref{eqn:delta_ns_final})--(\ref{eqn:nu}).    

In Sec.~\ref{sec:Grand}, we develop grand canonical formalism for the description of universal finite-$T$/finite-$L$ corrections 
to the equations of state and superfluid density. This approach is instructive from both technical and conceptual points of view. Technically, it allows to formulate the problem in terms of a single generating function expressed by Gaussian functional integral that can be readily calculated.  From the conceptual perspective, the form of this integral sheds a direct light on the origin of the universal $T^{d+1}$ and $1/L^{d+1}$ scaling of $\Delta n_s$ as well as of the corrections to the basic equations of state.

We validate the theory presented in Sec.~\ref{sec:Quant} by performing quantum Monte Carlo simulations for several representative 
system of interacting bosons on the square lattice in Sec.~\ref{sec:Num}. In Sec.~\ref{sec:disc}, we summarize our results and make concluding remarks.


\section{Finite-size effects in a classical U(1) system}
\label{sec:Class}

\subsection{General analysis}

Within the effective action formalism, the long-wave physics of a $d$-dimensional low-temperature quantum superfluid can be mapped onto that of a
finite-size (along the imaginary-time direction) $(d+1)$-dimensional classical counterpart. It is thus reasonable to put the problem in a somewhat broader context and first consider finite-size corrections
to the superfluid stiffness in a simple $d_c=(d+1)$ classical $XY$-type model with broken U(1) symmetry and equivalence between all directions.
Its effective long-wave action for the field of superfluid phase, 
$\Phi (r)$ is given by: 
\be\label{eqn:classical_action}
A[\Phi] = \int \,  {\rm d}^{d_c} r \left\{ 
\frac{\bar{\Lambda}_s}{2}(\nabla\Phi)^2+\frac{\theta}{4}[(\nabla \Phi)^2]^2 \right\} ,
\ee
(in the classical case, one can ``absorb" temperature into the definition
of coefficients and measure energies in units of temperature). Thermodynamically, parameter $\theta$ 
is defined through the second derivative of the free-energy density, ${\cal F}$, with respect to the square of the phase twist wavevector:
\be\label{eqn:sigma_classical}
\theta = 2\frac{\partial^2\mathcal{F}(k_0^2)}{\partial(k_0^2)^2}\bigg|_{k_0=0}\, .
\ee
In this sense, it is a direct analog of $\sigma$ in the quantum case, see Eq.~(\ref{eqn:nu}). 

The value of $\bar{\Lambda}_s$ in the effective theory (\ref{eqn:classical_action}) depends on the small but finite UV momentum cutoff $k_*$ so that $\bar{\Lambda}_s$ reaches its thermodynamic limit  value $\Lambda_s$ only in the $k_* \to 0$ limit. To the leading order, 
the difference between $\bar{\Lambda}_s$ and $\Lambda_s$ comes from the quartic term inducing fluctuational corrections to $\bar{\Lambda}_s$. These corrections are also responsible for the finite-size (Casimir-type) 
effects on the value of $\Lambda_s$. 

Two characteristic finite-size cases are particularly interesting. The first one is that of a hypercube sample when all linear system sizes are equal. The second one corresponds to a finite linear size, $L$, in one direction only. This case describes finite-temperature depletion in an infinitely large quantum superfluid. We will see that here the answer is reminiscent of (\ref{eqn:delta_ns_final}) with $\nu =  0$ and $c\beta$ replaced with $L$. An interesting fact that we demonstrate numerically is that the sign of $\theta$ can be both positive and negative, and can flip, as a function of control parameter(s), within one and the same simple statistical model. 

Apart from two special cases, there are reasons to consider a more general setup with linear system size $L_{\parallel}=L$ in one special (“longitudinal”) direction, and $L_{\perp}$ in all other directions. 
A convenient alternative parameterization of the sample geometry is through the aspect ratio:
\be\label{eqn:aspect}
\lambda=L/L_{\perp}.
\ee
The two special cases discussed above correspond to $\lambda=1 $ and $\lambda=0$. Other values of the aspect ratio provide a tool for
validation of the effective theory (\ref{eqn:classical_action}) and control over systematic errors in numeric data due to subleading corrections.   
   
The finite-size effects have both qualitative and quantitative aspects. At the qualitative level,  $\Lambda_s$ becomes anisotropic at any $\lambda \neq 1$: The longitudinal, $\Lambda_s^{(\|)}(L, \lambda)$, and transverse, $\Lambda_s^{(\perp)}(L, \lambda) $, values deviate from $\Lambda_s$ (and each other) due to different finite-size corrections: $\Delta \Lambda_s^{(\|)}(L, \lambda) \neq \Delta \Lambda_s^{(\perp)}(L, \lambda)$. 

Superfluid stiffness components, $\Lambda_s^{(x)}$,  
in an anisotropic system are defined though linear response 
to the twisted boundary condition applied in the $\hat{x}$-direction:
\be \label{eqn:n_s_def}
\Lambda_s^{(x)}=-\frac{TL_x ^2}{V}\frac{\partial^2\ln Z}{\partial\varphi_0^2}\bigg|_{\varphi_0=0}=-\frac{TL_x^2}{VZ}\frac{\partial^2Z}{\partial\varphi_0^2}\bigg|_{\varphi_0=0}\, .
\ee
Here $\varphi_0$ is the twist phase, $V$ is the system volume, 
and $Z$ is the partition function,
\be
Z = \int\mathcal{D}\Phi\, e^{-A[\Phi] }.
\label{part_func}
\ee
In this paper, we consider low-temperature superfluids in $d\geq 2$ and their classical counterparts in $d_c\geq 3$, when contributions from
phase-winding states are irrelevant for our analysis \cite{PS_2000_two_definitions}.

Imposing the twisted boundary condition on the field of phase, we have
\be
\Phi = k_0 x  +  \varphi\, , \qquad  k_0 = \varphi_0/L_x\, .
\ee
Since contributions from states with $2\pi $ windings in $\varphi$ are statistically negligible, terms linear in $\varphi_x \equiv \partial_x \varphi$ drop out from the action, and we have  
\begin{align}\label{eqn:A_xi}
\begin{split}
    \mathcal{A}(\nabla\Phi) & = \mathcal{A}(\nabla\varphi)+k_0^2\bigg[\frac{\bar{\Lambda}_s}{2}+\theta\varphi_x^2+\frac{\theta}{2}(\nabla\varphi)^2\bigg] \\
    & + k_0\theta(\nabla\varphi)^2\varphi_x+\mathcal{O}(k_0^3).
\end{split}
\end{align}
Terms mentioned in the second line of Eq.~(\ref{eqn:A_xi}) prove irrelevant and will be omitted. ${\cal O}(k_0^3)$ terms 
do not contribute to the second-order derivative (\ref{eqn:n_s_def}). 
Linear in $k_0$ term does produce a non-zero contribution to the
free energy but this contribution is proportional to the second 
power of $k_0$ and thus is subleading (as a $1/L^2$ correction)
to the contribution from quadratic in $k_0$ terms 
because it contains two extra derivatives.

From this point there are two ways to proceed: (i) Observing that the relevant terms in the action density (\ref{eqn:A_xi}) are quadratic, we can drop the rest of the terms and explicitly perform Gaussian integral (\ref{part_func}) thereby producing $Z$ and then applying (\ref{eqn:n_s_def}).  (ii) Taylor-expanding in powers of $k_0^2$ and then using (\ref{eqn:n_s_def}) before performing the Gaussian integral, we can express the result in terms of certain correlation functions. In this section, we employ approach (ii) allowing us to trace the correspondence with how we will treat the quantum system in the canonical ensemble in Sec.~\ref{sec:Quant}.  [Note also that approach (i) is a direct counterpart of grand canonical treatment of quantum system presented in Sec.~\ref{sec:Grand}.]

Expansion in $k_0^2$ yields the expression
\be
\Delta \Lambda_s^{(x)}=\frac{2\theta}{\Lambda_s}\Delta B +\frac{\theta}{\Lambda_s}\Delta C
\ee
for the depletion of the stiffness, where
\be
B=\Lambda_s\langle\varphi_x^2\rangle\,,\qquad C=\Lambda_s\langle \, (\nabla\varphi)^2\rangle\,.
\ee
Here we do not distinguish $\bar{\Lambda}_s$ and $\Lambda_s$
by assuming that all expressions are UV-regularized.
As opposed to the $ \langle \, (\nabla \varphi)^2 \rangle$ average 
which is insensitive to the direction of the twist, finite-size corrections to $ \langle  \varphi_x^2 \rangle$ depend on whether the $\hat{x}$-direction is longitudinal or transverse. 
While it appears that we have three averages to compute: 
$\Delta B_{\parallel}(L,\lambda)$, 
$\Delta B_{\perp}(L,\lambda)$ and 
$\Delta C(L,\lambda)$, they are related following identity
\be\label{eqn:corr_identity}
\Delta C = \Delta B_{\parallel}+(d_c-1)\Delta B_{\perp}\, .
\ee
In what follows, we will be expressing final answers in terms of 
$\Delta B_{\parallel}$  and $\Delta C$:
\be \label{eq:ns_par_gen}
\Delta \Lambda_s^{(\|)}\, =\, {\theta \over \Lambda_s} ( \Delta C  + 2 \Delta B_\|  ) \, ,
\ee
\bea 
\Delta \Lambda_s^{(\perp)} & =& {\theta \over \Lambda_s} ( \Delta C +  2 \Delta B_\perp ) \nonumber \\ & = &   {\theta \over (d_c-1) \Lambda_s} \, [ (d_c+1) \Delta C - 2 \Delta B_\| ]\, .
\label{eq:ns_perp_gen}
\eea
\subsection{Results}

\subsubsection{General case}

Being interested in the leading contributions, we calculate the averages based on the first (leading) term of the action (\ref{eqn:classical_action}).
Calculations are readily performed in the Fourier representation (the mode with zero wavevector is gauged out):
\be\label{eqn:fourier_gauge}
\varphi(\mathbf{r}) = \frac{\lambda^{d_c-1}}{L^{d_c}}\sum_{\mathbf{k}\neq0}\, \varphi_{\mathbf{k}}\, e^{i\mathbf{k}\cdot\mathbf{r}}\,,
\ee
\be\label{eqn:k_discrete}
\mathbf{k}\equiv\mathbf{k}_{\mathbf{n}}=\bigg(\frac{2\pi n_1}{L}, \,\frac{2\pi n_2}
{L_{\perp}},\,.\,.\,.\,,\,\frac{2\pi n_{d_c}}{L_{\perp}}\bigg)\,.
\ee
Here $\mathbf{n}=(n_1,\,n_2,\,.\,.\,.\,,\,n_d)$ is an integer vector; without loss of generality, direction ``1" is assumed to be the longitudinal direction.

In the Fourier representation, the average is given by 
\be\label{eqn:twopoint_classical}
\langle\, | \varphi ({\mathbf k})|^2\rangle=\frac{1}{n_sk^2}\,.
\ee
The next step is to write $B$ and $C$ as a sum over wavevectors. As discussed above, the effective action (\ref{eqn:classical_action}) implies a finite UV cutoff $k_*$, and hence the sums for $B$ and $C$ must be regularized by subtracting the corresponding integral with the same UV cutoff while replacing $\bar{\Lambda}_s$ with $\Lambda_s$. 
After regularization
\be\label{eqn:delta_C}
\Delta C = \frac{\lambda^{d_c-1}}{L^{d_c}}\sum_{\mathbf{k}\neq0}1-\int\frac{\mathrm{d}^{d_c} k}{(2\pi)^{d_c}}\,,
\ee
\be\label{eqn:delta_B}
\Delta B_{\parallel} = \frac{\lambda^{d_c-1}}{L^{d_c}}\sum_{\mathbf{k}\neq0}\frac{k_1^2}{k^2}
-\int\frac{\mathrm{d}^{d_c}k}{(2\pi)^{d_c}}\frac{k_1^2}{k^2}\,.
\ee

Next we observe that the difference between the sum and integral in the r.h.s of Eq.~(\ref{eqn:delta_C}) would yield identical zero if the sum contained the $\mathbf{k}=0$ term (by considering $1$ as the limiting case of a Gaussian with divergent width).  Thus, 
\be\label{eqn:C_class_fin}
\Delta C = -\frac{\lambda^{d_c-1}}{L^{d_c}}\,.
\ee
In anticipation of the final answer for $B_{\parallel}$, 
we define the $L$-independent quantity
\be
S_{\parallel}(\lambda)=L^{d_c}\Delta B_{\parallel}(L,\lambda)\,.
\ee
and use dimensionless vector
\be\label{eqn:q_def}
\mathbf{q}_{\mathbf{n}}=(n_1,\,\lambda n_2,\,.\,.\,.\,,\lambda n_{d_c}) \, ,
\ee
to express the answer as
\be \label{eq:S_d_lambda}
S_{\parallel} =\lambda^{d_c-1}\bigg[\sum_{\mathbf{n}\neq0}\frac{n_1^2}{\mathbf{q}_{\mathbf{n}}^2}-\frac{1}{d_c}\int\frac{\mathrm{d}^{d_c}q}{(2\pi)^{d_c}}\bigg]\,
\ee
(symmetrization of the integral in the r.h.s of (\ref{eqn:delta_B}) resulted in the prefactor $1/d_c$). 

This way Eqs.~(\ref{eq:ns_par_gen})--(\ref{eq:ns_perp_gen}) get converted in the following final answers:
\be
\Delta \Lambda_s^{(\|)}\, =\, {\theta/ \Lambda_s \over L^{d_c}}  \left[ 2S_{\parallel}(\lambda) - \lambda^{d_c-1}  \right]  ,
\label{n_s_long}
\ee
\be
\Delta \Lambda_s^{(\perp)}\, =\, -{\theta/ \Lambda_s \over (d_c -1) L^{d_c}}  \left[ 2S_{\parallel}(\lambda) + (d_c+1) \lambda^{d_c-1}  \right]  .
\label{n_s_tr}
\ee
Equations (\ref{n_s_long}) and (\ref{n_s_tr}) can be combined 
into relation
\be
\theta\, =\,  {(d_c -1)\Lambda_s L^{d_c}  
\over 2 [d_c S_{\parallel}(\lambda) + \lambda^{d_c-1}]}  
\left[  \Lambda_s^{(\|)} - \Lambda_s^{(\perp)} \right] 
\label{sigma_rel}
\ee
that proves very convenient for numeric validation of our theory
by comparing its r.h.s. to the value of $\theta$ computed from 
derivatives of the free energy.

\subsubsection{Hypercubic sample $(\lambda=1)$}

This case is particularly simple. Here we have $\Delta B_{\parallel}=\Delta B_{\perp}$, and from (\ref{eqn:corr_identity}) and (\ref{eqn:C_class_fin}) we immediately see that
\be
S_{\parallel}(\lambda=1)=-\frac{1}{d_c}
\ee
and
\be\label{eqn:ns_hypercube}
\Delta \Lambda_s(L) = -\frac{(1+2/d_c) \theta}{\Lambda_s L^{d_c}}\,.
\ee

\subsubsection{The $\lambda=0$ limit}

This case is of prime interest. We will not mention
$\lambda =0$ as an argument of functions throughout this section.
Now the sum over $n_2,\, n_3, \, \ldots , \, n_{d_c}$  in Eq.~(\ref{eq:S_d_lambda}) is replaced with the corresponding $(d_c - 1)$-dimensional integral and the remaining sum over $n_1$ can be performed analytically. In Sec.~\ref{subsec:math}, we consider this and similar sums that emerge in the finite-temperature and finite-size  expressions for $\Delta n_s$ in the quantum case. Here we utilize Eqs.~(\ref{alpha_4_def}), (\ref{f_4}), and
(\ref{alpha_4_res}) for $\alpha_4$ yielding
\be \label{eq:lambda_zero}
S_{\parallel}=-I_{d_c-1} \qquad (\lambda =0)\, ,
\ee
with $I_d$ defined by (\ref{eqn:Id_def}).
Hence,
\be\label{eqn:ns_long_classical}
\Delta \Lambda_s^{(\parallel)} = 
-2I_{d_c-1}\frac{\theta}{\Lambda_sL^{d_c}}\, ,
\ee
\be\label{eqn:ns_perp_classical}
\Delta \Lambda_s^{(\perp)}=\frac{2I_{d_c-1}}{d_c-1}\frac{\theta}{\Lambda_sL^{d_c}}\, ,
\ee
and
\be\label{eqn:sigma_rel}
\theta = \frac{(d_c-1)\Lambda_s L^{d_c} }{2d_c I_{d_c-1}}
\big[\Lambda_s^{(\perp)}(L) - \Lambda_s^{(\parallel)}(L)\big]\,.
\ee

\subsubsection{Sign of ``depletion"}

Since all corrections are proportional to $\theta$ their sign is controlled by the sign of $\theta$, which, as we will see later, can flip within one and the same model as a function of control parameter(s). The other two circumstances controlling the sign are: (i) the aspect ratio $\lambda$ and (ii) whether we are considering $\Delta \Lambda_s^{(\|)}$ or $\Delta \Lambda_s^{(\perp)}$. The hypercube case $\lambda = 1$ is particularly simple. Here $\Delta \Lambda_s^{(\|)} = \Delta \Lambda_s^{(\perp)} \equiv \Delta \Lambda_s$ and Eq.~(\ref{eqn:ns_hypercube}) tells us that the sign of depletion is always opposite to that of $\theta$. The same is generically true for $\Delta \Lambda_s^{(\|)}$, which is seen from Eq.~(\ref{eqn:ns_long_classical}). The case of $\Delta \Lambda_s^{(\perp)}$ is more interesting. By comparing the $\lambda=1$ and $\lambda=0$ results, Eqs.~(\ref{eqn:ns_hypercube}) and (\ref{eqn:ns_perp_classical}), respectively, we conclude that the sign of $\Delta \Lambda_s^{(\perp)}$ as necessarily changes within the $\lambda \in (0, 1)$ interval.

\subsection{Numeric simulations}

The goal of numeric simulations is two-fold. First, we want to validate our analytic approach by comparing to controlled results for a simple model. One possibility is to compare data for $\theta$ obtained by two distinctively different ways: (i) employing thermodynamic relation (\ref{eqn:sigma_classical}) and (ii) deducing the value of $\theta$ from the finite-size effect, Eq.~(\ref{eqn:sigma_rel}). As a by-product, 
we will also reveal the role of subleading corrections to the leading $\sim 1/L^{d_c}$ term. 
Our second goal is to demonstrate that the sign of $\theta$ can take
both positive and negative values and even flip---as a function of control parameter---within one and the same single-parametric model.

\subsubsection{The model}

The simplest model to simulate is the J-current model
\be
H_J/T \, =\,  {1\over 2K} \sum_b \, J_b^2 \, ,\qquad  J_b = -1,\, 0,\,  +1 \, .  
\label{J_current_model}
\ee
Here $b$ labels bonds of the $d_c$-dimensional hypercubic lattice, and $J_b$ is the bond current that takes on three integer values. Allowed values of the bond currents are also required to obey the zero divergence constraint on each site. The model has a simple Pollock-Ceperley type \cite{Pollock_Ceperley_relation} winding number estimator for $\theta$ based on Eq.~(\ref{eqn:sigma_classical}), while $K > 0$ is the only control parameter of the model. For the model to be in the ordered state, $K$ should be large enough. Our worm algorithm \cite{Prokofev2001} simulations of the ordered state were performed in $d_c=3$ dimensions at $K=0.4$ and $K=1.0$.  \\

\subsubsection{Winding number estimator for $\theta$}

Using 
\be
{\cal F} \,=\, - {1\over V} \ln Z  \qquad \qquad (T=1)\, ,
\label{F_density}
\ee
we have
\[
\theta \, =\, {2\over V} \left[   \left( {1\over Z}{\partial Z \over \partial (k_0^2)} \right)^2 -   
{1\over Z}{\partial^2 Z \over \partial (k_0^2)^2 } \right]_{k_0=0} \, .
\]
To relate this expression to the statistics of winding numbers (cf.~\cite{Pollock_Ceperley_relation}), we rewrite it in terms of the phase twist $\varphi_0$ in the direction $x$ [$k_0^2 =(\varphi_0/L_x)^2 $]:
\[
\theta \, =\, {2L_x^4 \over V} \left[   \left( {1\over Z}{\partial Z \over \partial (\varphi_0^2)} \right)^2 -   
{1\over Z}{\partial^2 Z \over \partial (\varphi_0^2)^2 } \right]_{\varphi_0=0}  \, .
\]
Expressing the full the partition function as the sum over 
winding numbers along the $x$-direction, $W_x$, 
\be
    Z \, =\,  \sum_{W_x} \, Z_{W_x}e^{i\varphi_0 W_x} \, =\,  \sum_{W_x} \, Z_{W_x} \cos (\varphi_0 W_x) \, ,
\ee
we see that
\[
    \frac{1}{Z} \left.~\frac{\partial Z}{\partial(\varphi_0^2)}\right|_{\varphi_0 = 0} = - \frac{1}{2Z}\sum_{W_x} Z_{W_x}W_x^2 \, =\,  - {1\over 2} \, \langle W_x^2 \rangle \, , 
\]
\[
    \frac{1}{Z} \left.~\frac{\partial^2 Z}{\partial(\varphi_0^2)^2}\right|_{\varphi_0 = 0} =  \frac{1}{12Z}\sum_{W_x} Z_{W_x}W_x^4 \, =\,   {1\over 12} \, \langle W_x^4 \rangle \, , 
\]
and thus
\be
\theta\, =\, {L_x^4 \over 6V} \left[ \, 3\,  \langle W_x^2 \rangle^2   -   \, \langle W_x^4 \rangle \right] \, .
\label{sigma_W}
\ee

\begin{figure}[tbh]
\includegraphics[width=1.0 \columnwidth]{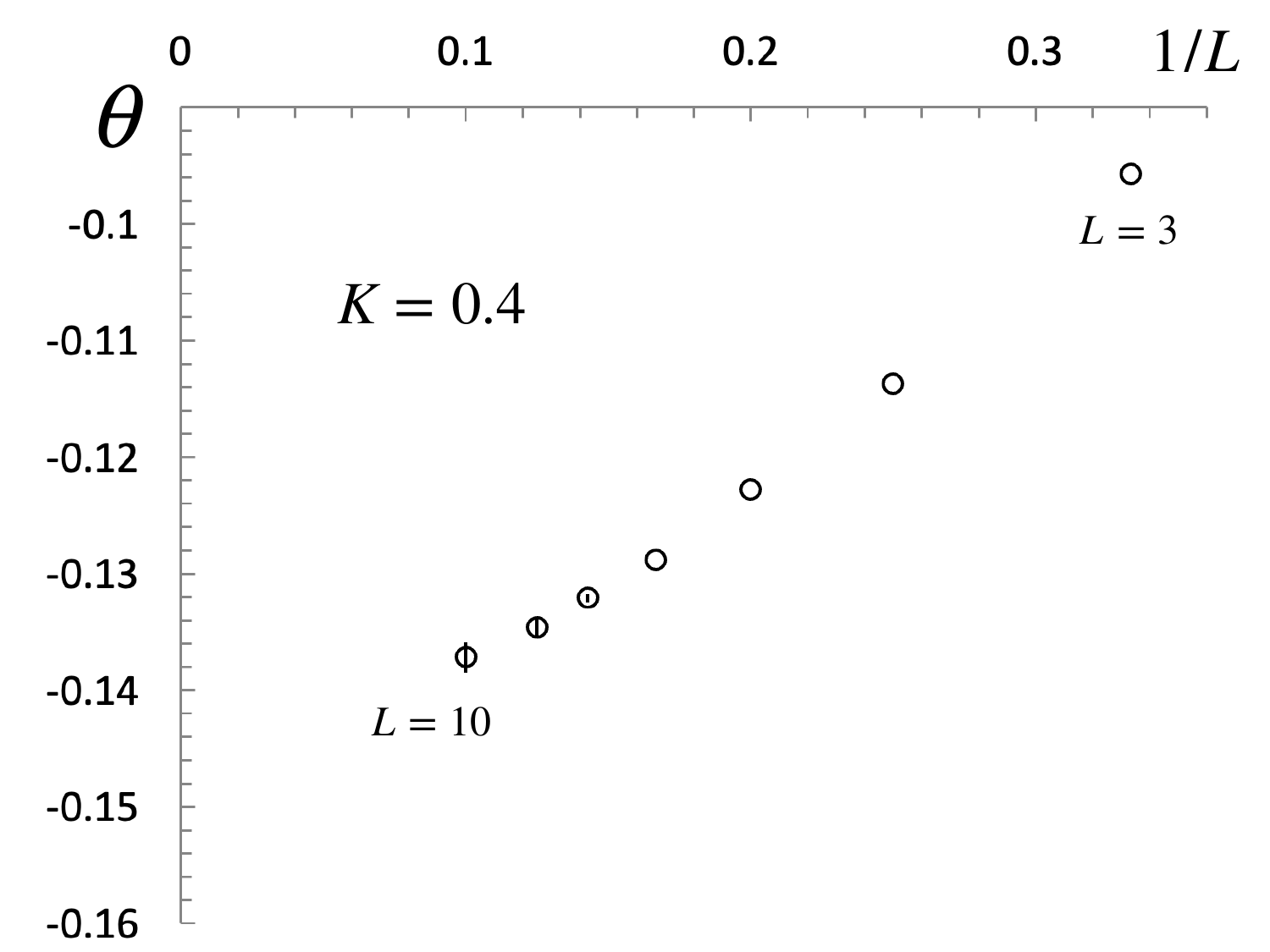}
\caption{Extracting parameter $\theta$ for the 3D J-current model (\ref{J_current_model}) at $K=0.4$ using estimator (\ref{sigma_W}) for hypercubic samples of different sizes. The error bars are shown only for $L=10,8$ and $7$; for $L < 7$, they are much smaller than the symbol size. Extrapolation to $L\to \infty$ yields $\theta=-0.14(1)$, which is consistent with the data for the $\Lambda_s$ depletion shown in Figs.~\ref{fig:n_s} and \ref{fig:Plot}.}
\label{fig:sigma}
\end{figure}

\begin{figure}[tbh]
\includegraphics[width=1.0 \columnwidth]{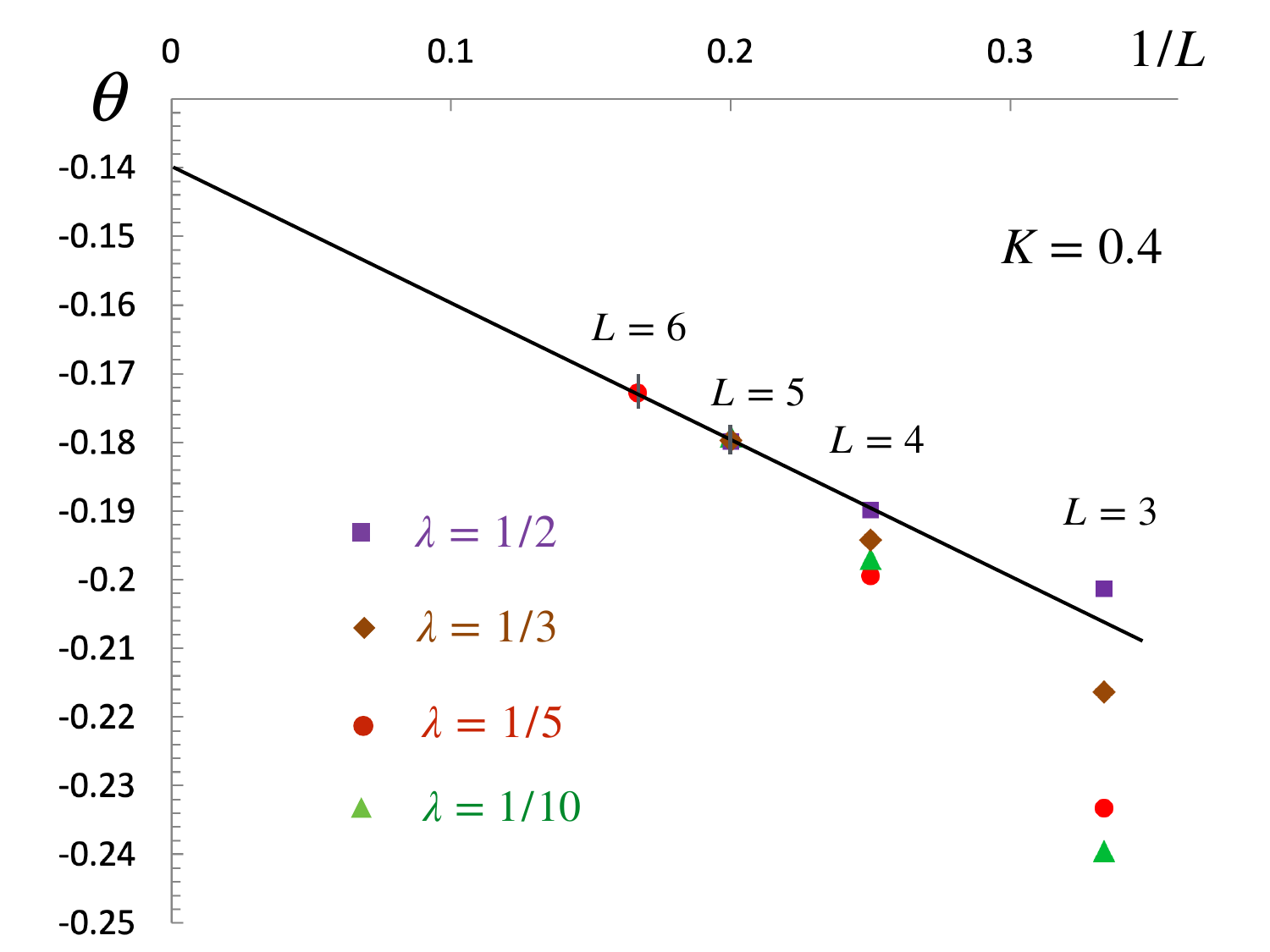}
\caption{Results for $\theta$ in the 3D J-current model (\ref{J_current_model}) at $K=0.4$ obtained for different system sizes and aspect rations using relation (\ref{sigma_rel}). The error bars for samples with $L<5$ are smaller than symbol size. Since at
$L=5$ the dependence on aspect ratio cannot be resolved within the error bars, simulation at $L=6$ was performed only for $\lambda=1/5$. The general trend is consistent with the thermodynamic limit reslt deduced from Fig.~(\ref{fig:sigma}) and suggests that the subleading correction scales as $1/L$ (solid line fit).}
\label{fig:Plot}
\end{figure}

\begin{figure}[tbh]
\includegraphics[width=0.9\columnwidth]{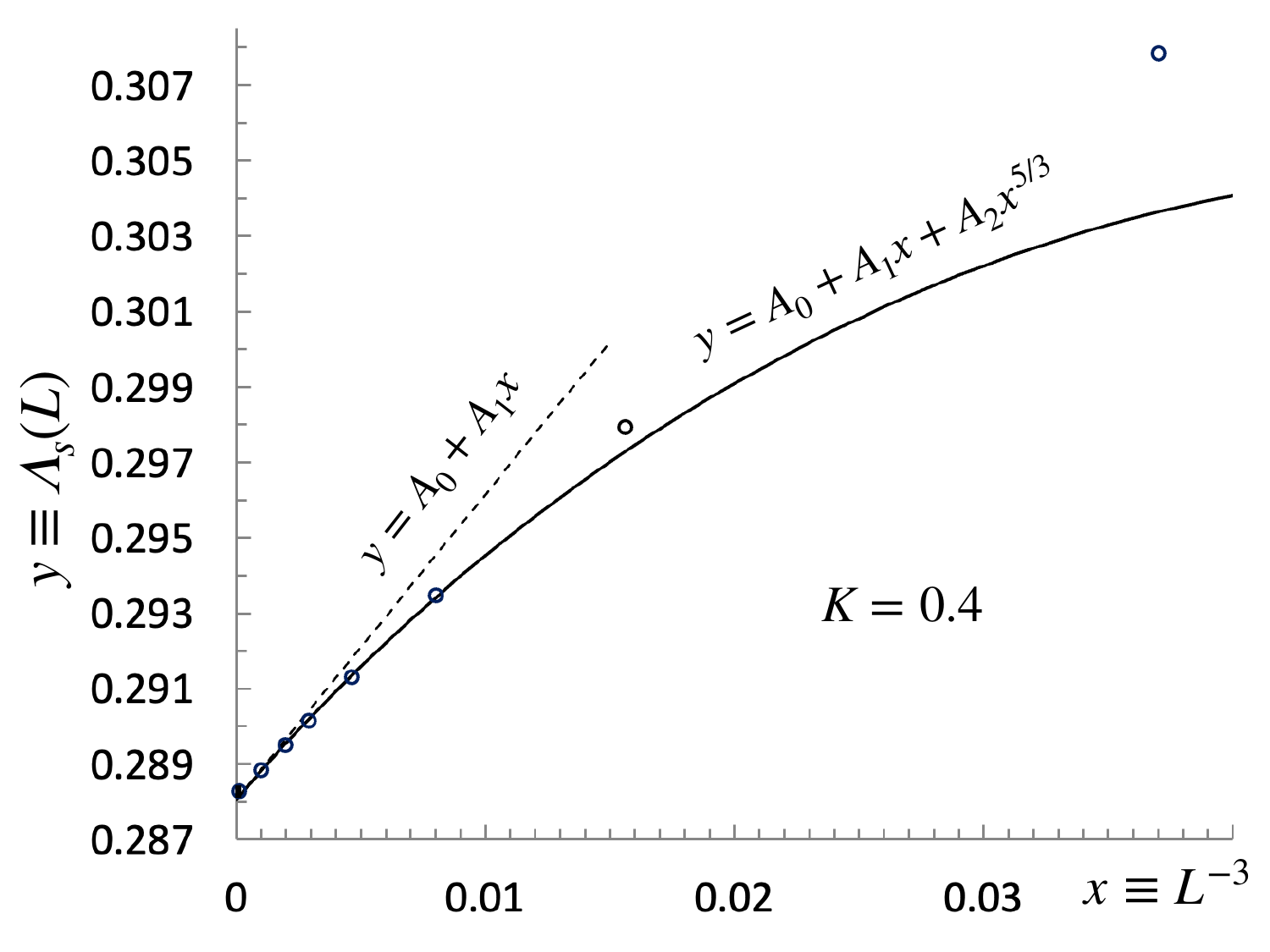}
\caption{Superfluid stiffness $\Lambda_s(L)$ of the 3D J-current model (\ref{J_current_model}) at $K=0.4$ for hypercubic samples with $L=3,\, 4,\, 5,\, 6,\, 7,\, 8,\,  10, \, 20$. 
The error bar is shown only for $L=20$; for other data points, the error bars are much smaller than the symbol size. We use the fitting ansatz $y(x)=A_0 + A_1 x + A_2 x^{5/3}$ (solid line) where the first two terms correspond to our theory and the subleading term is explained in the text. Parameters $A_0$ and $A_2$ ($A_0=0.28805$, $A_2=3.5$) are free fitting parameters while the value of $A_1$ is fixed by the relation (\ref{eqn:ns_hypercube}), implying that $A_1= -(5/3)\theta/\Lambda_s$, with $\Lambda_s \equiv A_0$, and independently measured $\theta=-0.14$,
see Figs.~\ref{fig:sigma} and \ref{fig:Plot}. 
The fit demonstrates perfect consistency with prediction of Eq.~(\ref{eqn:ns_hypercube}). It also illustrates---by comparison with the asymptotic $y(x)=A_0 + A_1 x$ law (dashed line)---that subleading contributions play significant role for system sizes $L\lesssim 10$.}
\label{fig:n_s}
\end{figure}

\begin{figure}[tbh]
\includegraphics[width=1.0 \columnwidth]{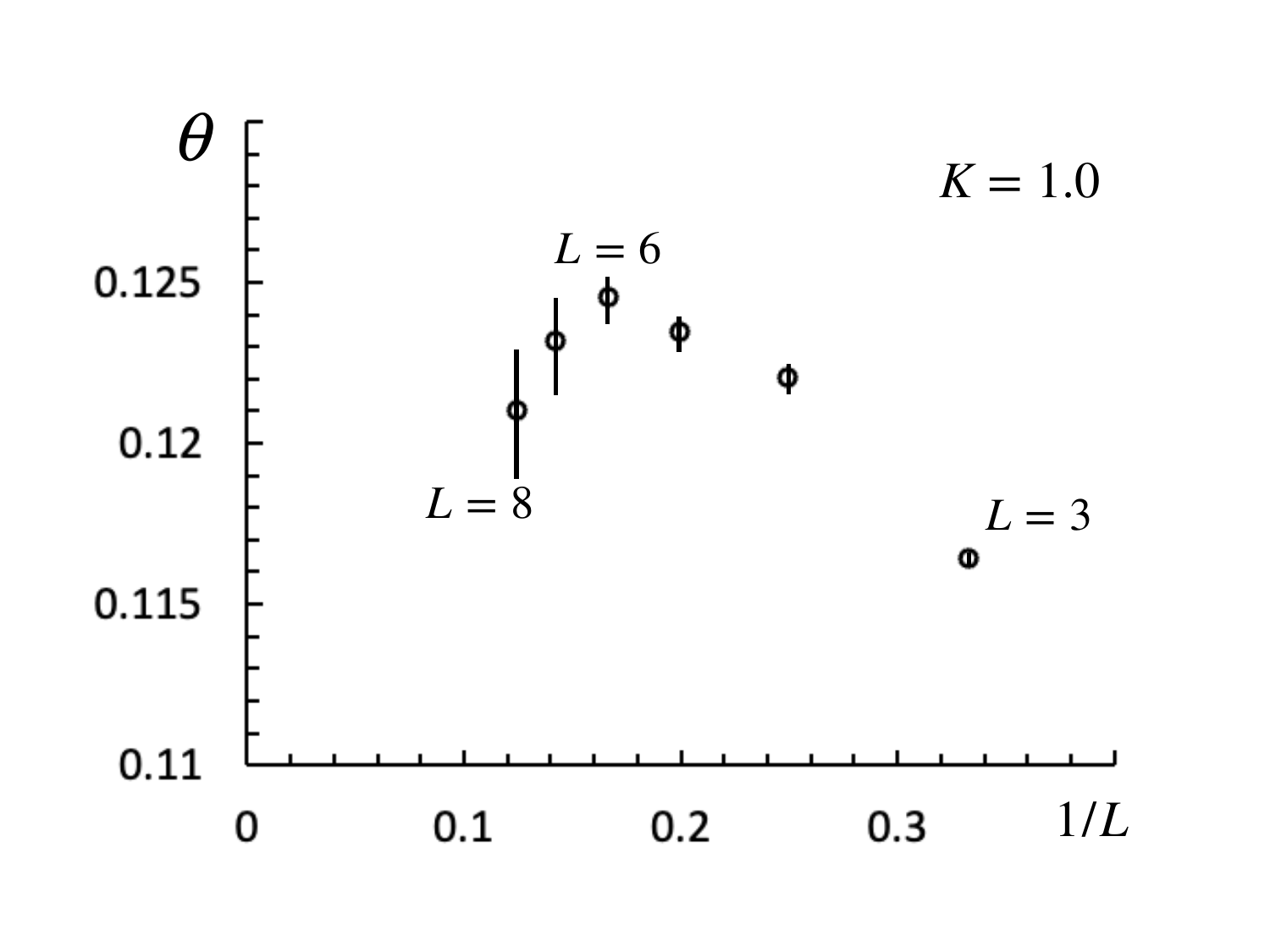}
\caption{Problem with extracting an accurate value of
$\theta$ for the 3D J-current model (\ref{J_current_model}) at $K=1.0$ using estimator (\ref{sigma_W}) for hypercubic samples. 
As opposed to the $K=0.4$ case (see Fig.~\ref{fig:sigma}), radical change of the dependence on system size at $L=6$
and rapidly increasing statistical errors for large $L$ 
prevents us from making a reliable extrapolation to the 
thermodynamic limit, which according to the data shown in Fig.~\ref{fig:n_s_1_0} is expected to be $\theta \approx 0.07$.
}
\label{fig:sigma_1_0}
\end{figure}

\begin{figure}[tbh]
\includegraphics[width=0.9\columnwidth]{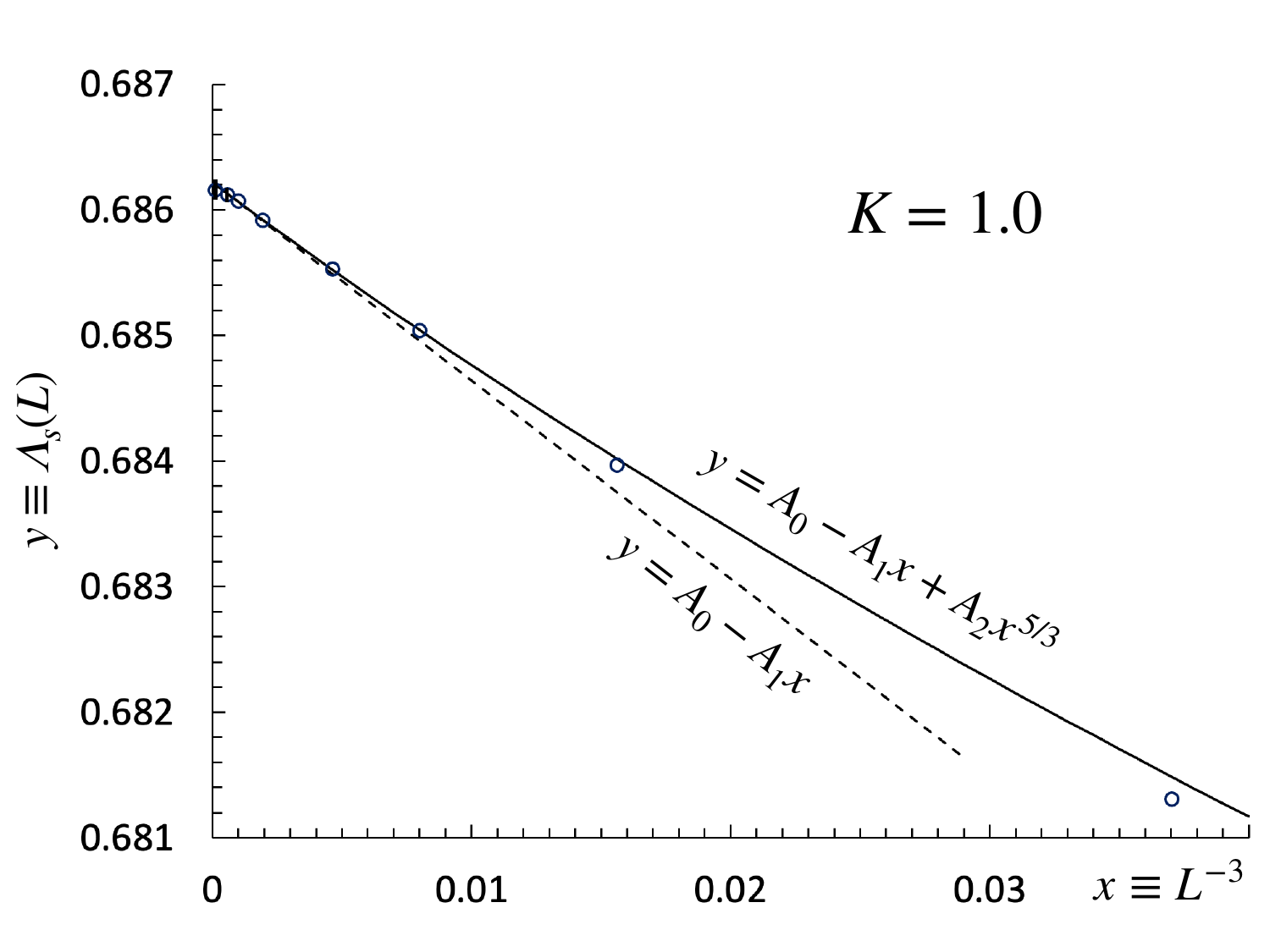}
\caption{Superfluid stiffness $\Lambda_s(L)$ for hypercubic samples of the 3D J-current model (\ref{J_current_model}) at $K=1.0$. 
The error bars are shown only for system sizes $L=20$ and $12$; 
they are much smaller than the symbol size for $L<12$.  Similar to Fig.~\ref{fig:n_s}, we use the fitting ansatz $y(x)=A_0 - A_1 x + A_2 x^{5/3}$ (solid line), but now with $\theta$  being a free fitting parameter in addition to $A_0$ and $A_2$;
the result is ($\theta=0.065$, $A_0=0.68622$, $A_2=0.27$).
In practice, $A_0$ is known very accurately from the largest system size simulation ($A_0 \approx \Lambda_s(L=20) = 0.6862$), and thus $A_1$ is directly related to $\theta$ by the relation $A_1= -(5/3)\theta/\Lambda_s$, see Eq.~(\ref{eqn:ns_hypercube}). 
The dashed line, $y(x)=A_0 -  A_1 x$, corresponds to the asymptotic linear behavior. }
\label{fig:n_s_1_0}
\end{figure}

\subsubsection{Numeric results}

Simulations of the model (\ref{J_current_model})  were performed for two values of $K$ corresponding (respectively)  to negative and positive sign of $\theta$: $K=0.4$ and $K=1.0$.

The $K=0.4$ data are presented in Figs.~\ref{fig:sigma}--\ref{fig:Plot}. Figure~\ref{fig:sigma} shows results for $\theta$ as a function of system size when it is computed using Eq.~(\ref{sigma_W}). 
From this plot we deduce that $\theta =-0.145(1)$ with large systematic uncertainty coming from ambiguity of how the data should be
extrapolated (see other examples and discussion below).

To verify Eqs.~(\ref{n_s_long}) and (\ref{n_s_tr}), we check their immediate implication---the prediction for $\theta$, Eq.~(\ref{sigma_rel})---as a function of system size and aspect ratio
by considering $\lambda \leq 1/2$. 
The values of $S_\parallel (\lambda)$ were calculated numerically using Eq.~(\ref{eq:S_d_lambda}) with matching UV cutoffs for the sum and the integral. We find that $S_\parallel (0.5) \approx -0.382$, which is very close to the $\lambda=0$ limit (\ref{eq:lambda_zero}), implying that for $\lambda \leq 1/3$, the difference between the exact result and (\ref{eq:lambda_zero}) is negligible.

Simulation results are presented in Fig.~\ref{fig:Plot}. For $L=3$, results for different values of $\lambda$'s substantially deviate from the limiting value $\theta=-0.14$, as well as from each other. 
These deviations are dramatically reduced for $L=4$, but remain pronounced. For $L=5$ we can no longer resolve the dependence on aspect ratio within the error bars.
Given that individual terms in (\ref{sigma_rel}) strongly depend
on the aspect ratio, absence of such dependence for larger $L$, 
confirms the validity of our considerations in the thermodynamic limit.
The data also suggest that the subleading correction to $\theta$ scales as $1/L$. Extrapolation of the ($\lambda=0.5$, $L=4,5,6$) set then leads to the estimate  $\theta = 0.14(1)$, which we now use to
validate the theory of depletion in the hypercube sample.

The data presented in Fig.~\ref{fig:n_s} demonstrate that
for the largest system sizes we correctly predict the leading
term (\ref{eqn:ns_hypercube}). Subleading contributions 
(we expect them to be proportional to $1/L^{(d_c+2)}$, see 
the text below Eq.~(\ref{eqn:A_xi}))
start playing a role for system sizes $L\lesssim 10$, forcing us to consider them when fitting the data over a broader range of system 
sizes, see Fig.~\ref{fig:n_s}.

The main goal of the simulation at $K=1.0$ is to reveal
the sign change of $\theta$ (and $\Delta \Lambda_s$) as a function of control parameter $K$.  The data presented in Figs.~\ref{fig:sigma_1_0} and \ref{fig:n_s_1_0}, clearly establish this fact. Somewhat unexpectedly (given system's behavior at $K=0.4$), here we face a problem of independently extracting an accurate value of $\theta$ from the winding number estimator (\ref{sigma_W}); see Fig.~\ref{fig:sigma_1_0}. The subleading corrections are not only 
strong but also non monotonic, dramatically changing the trend of finite-size effects at $L > 6$. Since large statistical errors 
prevent us from sampling larger system sizes, a reliable extrapolation to the thermodynamic limit is not possible. 
Nevertheless, the sign, the order-of-magnitude value, and the decreasing trend at $l \geq 7$ are qualitatively consistent with the value $\theta \approx 0.065$ value extracted from the fit in Fig.~\ref{fig:n_s_1_0}.

\section{The quantum system}
\label{sec:Quant}

Turning to the quantum system, we have two closely related but not identical options. From purely mathematical perspective, the most straightforward way to proceed is to start directly from Popov's hydrodynamic action, thereby immediately mapping the quantum $d$-dimensional problem onto the $(d+1)$-dimensional pseudo-classical U(1) model with complex-valued action. However, there is a subtlety. The Hydrodynamic action formalism works in the grand canonical ensemble while we are looking for the thermal depletion at a fixed density rather than than fixed chemical potential. That is why we prefer a somewhat different approach---a generalization of the approach previously used for the weakly interacting Bose gas \cite{book}---that starts with hydrodynamic Hamiltonian. 

\subsection{General analysis}\label{sec:q_gen_analysis}

The superfluid stiffness is the linear response coefficient relating supercurrent density (the expectation value of the persistent 
current density ${\bf j}$) to the (infinitesimal) value of ${\bf k}_0$:
\be
\langle\mathbf{j}\rangle=n_s\mathbf{k}_0 \qquad \qquad (k_0\rightarrow0)\, .
\ee
Current density in the long-wave limit is explicitly given by the derivative of the hydrodynamic Hamiltonian density over the phase gradient \cite{book}:
\be\label{eqn:current_def}
\mathbf{j} = \frac{\partial\mathcal{H}(\eta, \nabla\Phi)}{\partial(\nabla\Phi)}\,.
\ee
Here ${\cal H} (\eta, \nabla \Phi )$ is a function of 
$\nabla \Phi$ and $\eta$, with $\eta$ being the field of 
the number density fluctuations about the equilibrium expectation value $n$. By definition, ${\bf k}_0  = \langle \nabla \Phi \rangle$.
The hydrodynamic Hamiltonian is obtained by expanding the ground state energy density, ${\cal E}(n,k_0^2)$, in powers of $\eta$ and $k_0^2$, with subsequent substitution $k_0^2 \to  (\nabla \Phi)^2$ \cite{book}. Up to irrelevant for our purposes higher-order terms, the hydrodynamic Hamiltonian reads
\begin{align}\label{eqn:H_hydro_0}
\begin{split}
\mathcal{H}(\eta,\nabla\Phi) & = \frac{n_s^{(0)}}{2}(\nabla\Phi)^2+\frac{\eta^2}{2\varkappa}+\frac{\nu\eta}{2}(\nabla\Phi)^2 \\
& + \frac{\gamma\eta^2}{2}(\nabla\Phi)^2+\frac{\sigma}{4}(\nabla\Phi)^4\,.
\end{split}
\end{align}
Analogously to our approach in Sec.~\ref{sec:Class}, the main result of the quantum system, Eq.~(\ref{Delta_n_s_fin}), will be expressed in terms of correlation functions computed with respect to the Hamiltonian~(\ref{eqn:H_hydro_0}).
Compared to the simple classical effective action discussed in previous section however, the hydrodynamic Hamiltonian features several additional physical parameters; their meaning was described in the Introduction below Eqs.~(\ref{eqn:delta_ns_final})--(\ref{eqn:nu}). While $\sigma$ is the direct analogue of the parameter $\theta$ in the classical action~(\ref{eqn:classical_action}), the parameters $\nu$ and $\gamma$ are unique to the hydrodynamic Hamiltonian. According to Eqs.~(\ref{eqn:current_def}) and (\ref{eqn:H_hydro_0})
the current density is given by
\be\label{eqn:j_hydro}
\mathbf{j}=n_s^{(0)}\nabla\Phi + \nu\eta\nabla\Phi +\gamma\eta^2\nabla\Phi+\sigma(\nabla\Phi)^2\nabla\Phi\,.
\ee

To proceed, we separate the superflow phase gradient 
from phase fluctuations (without loss of generality we assume that $\mathbf{k}_0=k_0\hat{x}$),
\be\label{eqn:phi_decomp}
\Phi = k_0x+\varphi\, .
\ee
The bi-linear (in terms of density and phase fluctuations) 
Hamiltonian reads
\be\label{eqn:H_hydro}
\cH(\eta,\nabla\varphi)\, =\, \frac{n_s^{(0)}}{2}(\nabla\varphi)^2+\frac{\eta^2}{2\varkappa}+k_0\nu\eta\varphi_x\,.
\ee
As we will see, the higher-order terms prove irrelevant as long as we deal with the canonical ensemble.
Again, since $\varphi$ is periodic in the $x$-direction, we can 
safely omit the term $n_s^{(0)}\mathbf{k}_0\cdot\nabla\varphi$ from the Hamiltonian density (\ref{eqn:H_hydro}) because it has zero contribution to the action. 
In terms of our parameterization, the $x$-component 
of the current density operator is (omitting terms $\propto k_0^2$)
\bea
j_x  =  n_s^{(0)}k_0     &+& k_0[\gamma \eta^2 +  2 \sigma \varphi_x^2 + \sigma (\nabla \varphi)^2 ] + \nu \eta \varphi_x \nonumber \\
 &+&\,  n_s^{(0)} \varphi_x  \,+\, \nu k_0  \, \eta  \,  +\,  \sigma (\nabla \varphi)^2 \varphi_x \, .
\label{j_hydro_new}
\eea
All terms in the second line of Eq.~(\ref{j_hydro_new}) 
vanish upon averaging with the bi-linear Hamiltonian 
(\ref{eqn:H_hydro}) because they are based on odd powers of fluctuating fields. The omission of the second term in the second line is the key circumstance distinguishing the canonical ensemble from the grand canonical one. In the latter case, this term would generate two extra relevant contributions: one coming from the term $\propto \eta (\nabla \Phi)^2$ shown in the Hamiltonian (\ref{eqn:H_hydro_0}) and another one from the term $\propto \eta^3$ [not shown in (\ref{eqn:H_hydro_0}) in view of its irrelevance as long as we look for the depletion at constant density. The averages of the terms in the first line of Eq.~(\ref{j_hydro_new}) do not depend---in the leading order--on whether we calculate them in the canonical or grand canonical ensemble.

The temperature-induced change of the superfluid stiffness at a fixed number density thus comes as a sum of three terms 
(with appropriate UV regularization similarly to the classical case):
\be
\Delta n_s  = \Delta n_s^{(\nu)}  + \Delta n_s^{(\gamma)}  + \Delta n_s^{(\sigma)} \, ,
\label{three_terms}
\ee
\be
\Delta n_s^{(\nu)}  \, =\,   \, \nu \lim_{k_0\to 0}  \,{ \langle \eta  \,  \varphi_x \rangle \over k_0}
\, =\,  \nu \left. {\partial  \langle \eta  \,  \varphi_x \rangle \over \partial k_0} \right|_{k_0 = 0}\, ,
\label{D_nu}
\ee
\be
 \Delta n_s^{(\gamma)}  = \gamma \langle \eta^2 \rangle \, ,  \quad \Delta n_s^{(\sigma)} =  \sigma  \langle \,  (\nabla \varphi)^2 \rangle + 2\sigma  \langle  \varphi_x^2 \rangle \, .
\label{D_gamma_sigma}
\ee
As long as averages in Eqs.~(\ref{D_nu}) and (\ref{D_gamma_sigma}) do not vanish at $T=0$---in which case they have to be $k_*$-dependent---their $T=0$ contributions should be considered as quantum renormalization of the $k_*$-dependent $n_s^{(0)}$ value converting it to the genuine ground-state value of $n_s$, and, therefore, 
subtracted from the final answer for $\Delta n_s$. We will see that $\Delta n_s^{(\nu)}(\beta \to \infty) \to 0$, while it is clear 
that $ \Delta n_s^{(\gamma)} $ and $n_s^{(\sigma)}$ remain finite in the zero temperature limit. 

Evaluation of averages in Eqs.~(\ref{D_nu}) and  (\ref{D_gamma_sigma}) is done straightforwardly within the Popov's hydrodynamic action with the bi-linear Hamiltonian density \cite{Popov}  ($ \dot{\varphi}\equiv \partial_\tau \varphi$, $\beta =1/T$)
\be
S = \int_0^\beta d\tau {\cal L}(\eta, \dot{\varphi}, \nabla \varphi) \, , \qquad {\cal L}\, =\, -i\eta \dot{\varphi} - {\cal H} \, .
\label{Popov}
\ee
We illustrate the procedure by considering the average 
(\ref{D_nu}) as the most sophisticated one among the rest.
A convenient trick is to introduce the source term
\be
{\cal L}\, \to \, {\cal L}\, +\, \alpha \eta \varphi_x \, ,
\label{source}
\ee
in order to express the average in (\ref{D_nu}) through derivatives of the partition function logarithm. This way (\ref{D_nu}) reduces to ($V$ is the system's volume)
\be
\Delta n_s^{(\nu)} (\beta)   \, =\,   { \nu  \over \beta V} \left. {\partial^2 \ln Z \over \partial{k_0} \partial\alpha} \right|_{\alpha=k_0 = 0}\, ,
\label{depletion2}
\ee
\be
Z \, =\, \int e^{S[\eta, \varphi]} {\cal D}\eta  {\cal D}\varphi\, =\, {\rm const} \int e^{S_{\varphi} [\varphi]} {\cal D}\varphi\, .
\label{Z}
\ee
Action $S_{\varphi}[\varphi]$ is readily obtained by integrating out the Gaussian $\eta$-field; its Lagrangian density (up to irrelevant higher-order terms, and, in particular, terms $\propto \alpha^2$ and $\propto k_0^2$) is
\bea
{\cal L}_\varphi &=&  {\cal L}_\varphi^{(0)} \, +\, {\cal L}_\varphi^{(1)} \, , \\
\label{L_phi}
 {\cal L}_\varphi^{(0)}  &=&  - {\varkappa\over 2}  \dot{\varphi}^2 - {n_s^{(0)}\over 2} (\nabla \varphi)^2   \,  ,\\
\label{L_phi_0}
{\cal L}_\varphi^{(1)} &=&  i\left(  \nu k_0  -  \alpha\right) \varkappa \dot{\varphi} \varphi_x
\, -\, \varkappa \nu k_0 \alpha  \varphi_x^2 \, .~\label{L_phi_1}
\eea
This brings us to the following expression
\be
\Delta n_s^{(\nu)}\, =\,  \nu^2  \left [  \varkappa^2  \!\! \int \! d\tau  d^d r \, {\cal K}({\bf r}, \tau)  \, -\, \varkappa   \, \langle \varphi_x^2 \rangle \right] ,
\label{main}
\ee
\be
{\cal K}({\bf r}, \tau) \, =\,  \langle \, \dot{\varphi}({\bf r}, \tau)  \, \varphi_x ({\bf r}, \tau) \, \dot{\varphi}(0,0) \,  \varphi_x (0,0)\,  \rangle_0 \, ,
\label{K}
\ee
where the average is taken with respect to the action 
$S_0 = \iint d\tau d^dr {\cal L}_\varphi^{(0)}$.
Evaluation of the correlator ${\cal K}({\bf r}, \tau)$ is straightforward due to the Gaussian character of $S_0$.
The first step is applying Wick's theorem with the observation that $\langle \, \varphi_x ({\bf r}, \tau) \, \dot{\varphi}({\bf r}, \tau)\,  \rangle = \langle \, \varphi_x (0, 0) \, \dot{\varphi}(0, 0)\,  \rangle = 0$ by $x\to -x$ and  $\tau \to -\tau$ symmetries.
This yields 
\bea
{\cal K}({\bf r}, \tau) & =& \langle \, \dot{\varphi}({\bf r}, \tau) \, \dot{\varphi}(0,0)\,  \rangle_0 \, \langle  \, \varphi_x ({\bf r}, \tau)  \,  \varphi_x (0,0)\,  \rangle_0 \nonumber \\
 &+& \langle \, \dot{\varphi}({\bf r}, \tau)\,  \varphi_x (0,0) \,  \rangle_0 \, \langle  \, \varphi_x ({\bf r}, \tau)  \, \dot{\varphi}(0,0) \,  \rangle_0 \, ,~\qquad
\label{K_reduced}
\eea
thereby reducing the problem to the product of two standard phase-phase correlators $\langle \,  \varphi ({\bf r}, \tau) \,  \varphi (0,0)\,  \rangle_0 $ differentiated with respect to either $\tau$ and/or $x$.

Thus, the answer for $\Delta n_s$ comes in the form of finite-$\beta$ corrections, $\Delta A$,  $\Delta B$,  $\Delta C$ and $\Delta D$ to the correlators
\be
A= \varkappa^2 \!  \! \int \! \! d\tau  d^d r \, {\cal K}({\bf r}, \tau)\, ,
\label{A}
\ee
\be
 B=  \varkappa \, \langle  \varphi_x^2 \rangle \, , \quad  C=  \varkappa \, \langle \, (\nabla  \varphi)^2 \rangle \, , \quad D=\frac{\varkappa}{c^2} \, \langle\dot{\varphi}^2 \rangle\, ,
\label{BCD}
\ee
where we have taken into account $\eta=-i\varkappa\dot{\varphi}$. Specifically, we have
\be
\Delta n_s \, =\, \nu^2 (\Delta A -\Delta B) +  {2 \sigma \over \varkappa} \Delta B +  \frac{\sigma}{\varkappa} \Delta C - \gamma n_s \Delta D
\, .
\label{Delta_n_s_gen}
\ee

One can literally ``read off the page" the phase-phase correlator in 
Fourier space by looking at Eq.~(\ref{L_phi}):  
\be
\langle|\varphi_{\mathbf{k},\omega}|^2\rangle=\frac{1}{\varkappa\omega^2+n_sk^2}\, .
\ee
Here $\omega$ stands for bosonic Matsubara frequency. With rescaling by the velocity of sound, $c \beta$ plays the role of an extra spatial dimension in the harmonic action. Hence, it is convenient to introduce the ``imaginary-time wavevector"
\be\label{eqn:zeta_def}
\zeta \equiv \omega/c\, .
\ee
What is left is evaluation of explicit expressions for 
$A$, $B$, $C$ and $D$: 
\be\label{eqn:A-corr}
A = \frac{2}{c^2\beta V}\sum_{\mathbf{k}, \zeta}\frac{k_x^2\zeta^2}{(\zeta^2+k^2)^2}\,,
\ee
\be\label{eqn:BC-corr}
B = \frac{1}{c^2\beta V}\sum_{\mathbf{k},\zeta}\frac{k_x^2}{\zeta^2+k^2}\,, \quad C = \frac{1}{c^2\beta V}\sum_{\mathbf{k},\zeta}\frac{k^2}{\zeta^2+k^2}\,,
\ee
\be\label{eqn:D-corr}
D = \frac{1}{c^2\beta V}\sum_{\mathbf{k}, \zeta}\frac{\zeta^2}{\zeta^2+k^2}\,.
\ee

The structure of Eq.~(\ref{Delta_n_s_gen}) reveals 
similarities with the classical system. Calculating the finite-temperature correction to $n_s$ is completely analogous to computing the finite-size correction (by discretization of the corresponding wavevector values). Thus, the problem is mathematically equivalent to that of an anisotropic classical system, this time with \textit{three} inequivalent directions.

Similar to the calculations in Sec.~\ref{sec:Class}, we will assume that the $\mathbf{k},\omega = 0$ mode has been subtracted from the Fourier expansion of the field $\varphi(\mathbf{r},\tau)$. With this particular choice of ``gauge," we can write $\Delta D$ in terms of $\Delta C$ as 
\be\label{eqn:D_C_relation}
\Delta D = -\frac{1}{c^2\beta V}-\Delta C\,,
\ee
reducing the number of correlators to be calculated to three. Note that the first term on the right hand side of Eq.~(\ref{eqn:D_C_relation}) vanishes if at least one of the linear system sizes is taken to infinity, or if temperature is set to zero.

The calculation of the correlators $A$, $B$ and $C$ involves explicitly evaluating certain standard sums and integrals. The problem reduces to evaluating four coefficients $\alpha_1,\dots\alpha_4$. We refer to Appendix \ref{subsec:math} for the mathematical details, and will in the rest of this section directly present results of the calculations for certain cases of interest.

\subsection{Finite-temperature corrections}

This case is of main interest because it directly brings us to the generalization of Landau formula (\ref{eqn:delta_ns_final}) for 
the superfluid stiffness depletion. For finite $\beta$ and infinite spatial directions, the only discrete wavevector in (\ref{eqn:A-corr})--(\ref{eqn:BC-corr}) is $\zeta$. After symmetrization of the integrand and replacement $k_x^2$ with $k^2/d$, we introduce a dimensionless wavevector $\mathbf{q}$
\be
\mathbf{q} = c\beta \mathbf{k}\,,
\ee
and observe, using Eqs.~(\ref{alpha_1_def}), (\ref{alpha_1_res}), (\ref{alpha_3_def}) and (\ref{alpha_3_res}), that
\begin{align}
c(c\beta)^{d+1}\Delta A(\beta) & = \frac{2}{d}\alpha_3=-I_d,\label{eqn:A_beta_result} \\
c(c\beta)^{d+1}\Delta B(\beta) & = \frac{1}{d}\alpha_1=\frac{I_d}{d}, \label{eqn:B_beta_result}\\
c(c\beta)^{d+1}\Delta C(\beta) & = \alpha_1=I_d.
\end{align}
Hence, the finite-$\beta$ depletion of $n_s$ is given by
\be
{c(\beta c)^{d+1}\Delta n_s (\beta)\over I_d}  \, =\, -{d\! +\! 1\over d} \, \nu^2 \,+\, {d\! +\! 2\over d} \, {\sigma\over \varkappa}
 \,+\, \gamma n_s \, .
\label{Delta_n_s_fin}
\ee

\subsection{Finite-size corrections}

Assume now that the linear system size in one of the directions, call it the $\hat{f}$-direction, be finite and equal to $L$ (with the periodic boundary condition implied), while in other directions systems sizes and $\beta$ are infinite. In this situation $n_s$ becomes anisotropic: the longitudinal 
(along $\hat{f}$-direction), and transverse values of the superfluid stiffness are now different due to fine-size corrections: $\Delta n_s^{(\|)}(L) \neq \Delta n_s^{(\perp)}(L)$. As in the classical system, the phase fluctuations depend on whether the directions
$\hat{f}$ and $\hat{x}$ are parallel or orthogonal: 
$\Delta A_\|(L) \neq \Delta A_\perp(L)$ and $\Delta B_\|(L) \neq \Delta B_\perp(L)$. [Note that $\Delta C(L)$ remains isotropic
by the spatial symmetry of the sum in Eq.~(\ref{eqn:BC-corr}).] 

The sum for $\Delta A_\|(L) $ is identical to that for 
$\Delta A(\beta) $ in an infinite system---up to swapping
$\tau$ and $x$ variables places and rescaling by the sound velocity. 
A direct analog of (\ref{eqn:A_beta_result}) reads:
\be
cL^{d+1}\Delta A_\|(L) \, =\, {2 \over d} \, \alpha_3 \, =\, -I_d \, .
\label{A_long}
\ee
Exactly the same considerations apply to $\Delta B_\perp(L)$:  
\be
cL^{d+1}\Delta B_\perp(L) \, =\, {1\over d}\, \alpha_1 \, =\, {I_d \over d}  \, .
\label{B_trans}
\ee

Sums for $\Delta A_\perp(L)$ and $\Delta B_\|(L)$ do not have finite-$T$ analogs.  In the case of $\Delta B_\|(L)$, we substitute
\be
k_x \, =\, 2 \pi n/L \, , \qquad n=0, \pm 1, \pm 2, \ldots \, ,
\label{param_1_1}
\ee
in to Eq.~(\ref{eqn:BC-corr}) and observe that the calculation reduces to that of the $\alpha_4$ coefficient in Appendix \ref{subsec:math}. Using Eqs.~(\ref{alpha_4_def}) and (\ref{alpha_4_res}), we have
\be
cL^{d+1}\Delta B_\|(L) \, =\, \alpha_4 \, =\, -I_d \, ,
\label{B_long}
\ee
Likewise, to calculate $\Delta A_\perp(L)$, we substitute 
\be
k_f \, =\, 2 \pi n/L \, , \qquad n=0, \pm 1, \pm 2, \ldots \, ,
\label{param_2_1}
\ee
in to Eq.~(\ref{eqn:A-corr}) to map the calculation to that of the $\alpha_2$ coefficient in Appendix \ref{subsec:math}, resulting in
 \be
cL^{d+1}\Delta A_\perp(L) \, =\, {2\over d(d+2)} \, \alpha_2 \, =\, {I_d\over d} \, .
\label{A_trans}
\ee
Note that in integrals over the space of ${\bf q}=(q_1, q_2, q_3, \ldots, q_d)$ a factor $q_1^2 q_2^2 $ can be converted into  $q^4/[d(d+2)]$ upon hyper-angle averaging, see Appendix \ref{appendix_symm}.
Finally, since isotropic $\Delta C$ is trivially decomposed into 
B-sums
\be
\Delta C = (d-1)B_\perp(L)   + B_\|(L) \, ,
\label{C_trans_long}
\ee
we have
\be
cL^{d+1} \Delta C = -  {I_d \over d}  \, .
\label{C_trans_long}
\ee

This brings us to the results
\be
{cL^{d+1}\Delta n_s^{(\|)}(L) \over I_d}  \, =\, - {(2d\!+\! 1)\sigma \over d \varkappa} \,-\,  {\gamma n_s\over d}  \, ,
\label{Delta_n_s_long}
\ee
\be
{cL^{d+1}\Delta n_s^{(\perp)}(L) \over I_d}  \, =\, {\sigma \over d \varkappa} \,-\,  {\gamma n_s\over d}  \, .
\label{Delta_n_s_trans}
\ee

Independence of finite-size corrections on $\nu =dn_s/dn$ might
appear surprising. Nevertheless, this outcome can be immediately deduced---by contradiction---from Landau theory of superfluidity in a Galilean system. For such a system, $\gamma = \sigma =0$,
while  $\nu=1/m \neq 0$, and existence of a $\nu$-dependent finite-size correction would imply dependence of ground-state stiffness on $L$, in contradiction with the fact that in a Galilean system
$n_s(T=0)\equiv n$, no matter whether the system size is infinite or finite.

Similarly to Eq.~(\ref{eqn:sigma_rel}), an expression for $\sigma$ in terms of $\Delta n_s^{(\parallel)}$ and $\Delta n_s^{(\perp)}$ can be obtained by subtracting Eq.~(\ref{Delta_n_s_long}) from Eq.~(\ref{Delta_n_s_trans}), resulting in
\be\label{eqn:sigma_q_anis}
\sigma = \frac{d\varkappa c L^{d+1}}{2I_d(d+1)}(n_s^{(\perp)}-n_s^{(\parallel)}).
\ee
This expression serves as a useful numeric check against Eq.~(\ref{sigma_wind}), since $n_s^{(\parallel)}$ and $n_s^{(\perp)}$ can easily be obtained from simulations of anisotropic systems by sampling winding numbers in different spatial directions.

\subsection{Finite temperature and system size}\label{sec:fin_T_L}
In numeric simulations, both $c\beta$ and linear system size(s) are finite and not necessarily equal to each other. Let us consider a typical case of a spatial hypercube with linear size $L$. Following treatment similar to that in Sec.~\ref{sec:Class}, we define the dimensionless ratio
\begin{equation}
    \lambda_{\beta} = \frac{c\beta}{L},
\end{equation}
so that all results can be expressed as functions of
$(\beta, \lambda_{\beta})$. 

Now all wavevectors are discrete and given by
\begin{equation}
    \zeta = \frac{2\pi n_0}{c\beta}, \qquad k_i = \frac{2\pi n_i}{L} \quad (i\neq 0),
\end{equation}
where $n_0$ and $n_i$ are integers. Defining the $(d+1)$-dimensional vectors
\begin{equation}
    \mathbf{n} = (n_0, n_1, n_2, \dots, n_d)
\end{equation}
and
\begin{equation}
    \mathbf{q_n} = (\lambda_{\beta}^{-1}n_0, n_1, \dots, n_d),
\end{equation}
the regularized sums for $\Delta A$, $\Delta B$, and $\Delta C$, upon symmetrization of the integrals, take the form
\begin{equation}\label{eqn:A_nondim}
    \begin{split}
    \Delta A = \frac{\lambda_{\beta}^d}{c(c\beta)^{d+1}}\bigg(&\lambda_{\beta}^{-2}\sum_{\mathbf{n}}\frac{n_0^2n_1^2}{\mathbf{q}_{\mathbf{n}}^4} \\ 
    & -\frac{1}{(d+1)(d+3)}\int\mathrm{d}^{(d+1)}q\bigg),
    \end{split}
\end{equation}
\begin{equation}\label{eqn:B_nondim}
    \Delta B = \frac{\lambda_{\beta}^d}{c(c\beta)^{d+1}}\bigg(\sum_{\mathbf{n}}\frac{n_1^2}{\mathbf{q}_{\mathbf{n}}^2}-\frac{1}{d+1}\int\mathrm{d}^{(d+1)}q\bigg),
\end{equation}
\begin{equation}\label{eqn:C_nondim}
    \begin{split}
        \Delta C = d\Delta B,
    \end{split}
\end{equation}
\begin{equation}\label{eqn:D_nondim}
    \Delta D = -\frac{\lambda_{\beta}^d}{c(c\beta)^{d+1}}-d\Delta B
\end{equation}

Eq.~(\ref{eqn:C_nondim}) immediately follows from Eq.~(\ref{eqn:BC-corr}) by isotropy in spatial directions, while Eq.~(\ref{eqn:D_nondim}) follows from combining Eq.~(\ref{eqn:D_C_relation}) and Eq.~(\ref{eqn:C_nondim}). For details of the symmetrization procedure leading to the prefactor in front of the integral appearing in Eq.~(\ref{eqn:A_nondim}), see Appendix~\ref{appendix_symm}.

\subsection{Spacetime hypercube at $\nu=0$}
Of special interest and importance for our numeric simulations (see below) is the case of spacetime hypercube ($c\beta = L$) at $\nu=0$, which is quite similar to the $\lambda=1$ case in the classical model (\ref{eqn:classical_action}).

At $\nu=0$ there is no contribution from $\Delta A$. 
The remaining terms $\Delta B$ and $\Delta C$ are easily mapped to those appearing in the calculation for the classical system. Thus, we end up with the formula
\begin{equation}\label{st_hypercube}
    c(c\beta)^{d+1}\Delta n_s = \frac{\gamma n_s}{d+1}-\frac{(2\!+\!d)\sigma}{(d\!+\!1)\varkappa} \qquad (c\beta = L ,\,  \nu=0)\, .~~
\end{equation}
A notable feature of Eq.~(\ref{st_hypercube}) is that if the parameter $\sigma$ is negative and large enough in modulus, it is possible to observe an increase of the superfluid stiffness as temperature is increased and the system size is decreased, while keeping $c\beta = L$. 

\section{Grand Canonical Formalism}
\label{sec:Grand}

A key feature of the Hydrodynamic Hamiltonian approach is that it (as described in Sec.~\ref{sec:q_gen_analysis}) allows us to work in the canonical ensemble by systematically discarding contributions to $\Delta n_s(n,T)$ originating from temperature-induced change in the total density at a fixed chemical potential. In many experimental contexts however, the total density is not held constant in measurements of superfluid density. In such cases, we must work in the grand canonical ensemble, and instead calculate the depletion $\Delta n_s(\mu,T)$ at constant chemical potential.

Apart from some special cases (for instance systems with particle-hole symmetry), the depletion of superfluid stiffness in the two ensembles is not the same, $\Delta n_s(\mu,T) \neq\Delta n_s(n,T)$. The two are however not completely independent; they are related to each other through the equation of state $\Delta n(\mu,T)$ as
\begin{equation}\label{eqn:ns_n_relation}
    \Delta n_s(n,T) = \Delta n_s(\mu,T)-\nu \Delta n(\mu,T) \, .
\end{equation}

It might appear that the grand-canonical treatment of $n_s(n,T)$ involves a substantial extra price of evaluating $\Delta n(\mu,T)$ in addition to $\Delta n_s(\mu,T)$.
As we will see below, this is not the case thanks to significant technical advantages. The key aspect here is that both $\Delta n_s(\mu,T)$ and $\Delta n(\mu,T)$ can be readily obtained---along with finite-$T$/finite-$L$ corrections to all other basic thermodynamic quantities---within a unified approach based on a single generating function. The generating function is expressed as Gaussian functional integral that can be readily calculated without explicitly resorting to correlators. Furthermore, now it is sufficient to work  with the phase-only representation of Popov's hydrodynamic action.


\subsection{Effective Lagrangian}

The effective Lagrangian in the phase-only representation can be obtained by Taylor-expanding the broadly-understood---as the negative density of the grand canonical potential---ground-state pressure, $p_0(\mu, k_0^2)$, in powers of the small variation of the chemical potential, $\delta \mu$, and square of superflow wavevector $k_0^2$, with the subsequent replacement $\delta \mu\rightarrow -i\dot{\Phi}$ and $k_0^2\rightarrow(\nabla\Phi)^2$ \cite{book}. The expansion coefficients of this representation can be related to the ones appearing in the hydrodynamic Hamiltonian, Eq.~(\ref{eqn:H_hydro_0}), see Appendix~\ref{appendix:action}. The effective Lagrangian density with irrelevant terms omitted then takes the form
\begin{equation}\label{eqn:phase_lagrangian}
    \begin{split}
    \mathcal{L}(\Phi) = &-\frac{n_s^{(0)}}{2}(\nabla\Phi)^2-\frac{\varkappa}{2}\dot{\Phi}^2-\frac{i\varkappa\nu}{2}\dot{\Phi}(\nabla\Phi)^2 \\
    &+\frac{\varkappa^2\tilde{\gamma}}{2}\dot{\Phi}^2(\nabla\Phi)^2-\frac{\tilde{\sigma}}{4}(\nabla\Phi)^4\,,
    \end{split}
\end{equation}
where
\begin{equation}
    \tilde{\gamma}=\gamma-\frac{\varkappa\nu\lambda}{2},\qquad\tilde{\sigma}=\sigma-\frac{\varkappa\nu^2}{2}\,.
\end{equation}
The only parameter here not appearing in the hydrodynamic Hamiltonian is 
\be\label{eqn:lambda_derivative}
\lambda \, =\, \frac{\partial^3 \mathcal{E}(n,k_0^2)}{\partial n^3}\bigg|_{k_0=0} \, =\, 
{d^2\mu \over dn^2} \, =\, -{1\over \varkappa^3} {d \varkappa \over d\mu} \, .
\ee

\subsection{Generating function}

Our main task is to calculate the grand potential density
\begin{equation}
    \frac{1}{V}\Omega(\mu, T, k_0) = -\frac{T}{V}\ln Z(\mu, T, k_0)\,,
\end{equation}
which acts as a generating function for the rest of thermodynamic quantities. In particular, the total density is given by the standard thermodynamic relation
\begin{equation}\label{eqn:n_derivative}
    n(\mu, T) = -\frac{1}{V}\frac{\partial\Omega(\mu, T, k_0=0)}{\partial\mu} \, ,
\end{equation}
while the superfluid stiffness $n_s(\mu, T)$ is given by the second derivative
\begin{equation}\label{eqn:ns_derivative}
    n_s(\mu, T) = \frac{1}{V}\frac{\partial^2\Omega(\mu, T, k_0)}{\partial k_0^2}\bigg|_{k_0=0}\,.
\end{equation}
As before, $k_0$ is the wave vector of the infinitesimal superflow. We thus proceed similarly. Decomposing $\Phi$ as in Eq.~(\ref{eqn:phi_decomp}) and inserting it into Eq.~(\ref{eqn:phase_lagrangian}) yields
\begin{equation}\label{eqn:L_renormalized}
    \begin{split}
    \mathcal{L}'(\varphi) =&-\frac{n_s^{(0)}+\tilde{\sigma}k_0^2}{2}(\nabla\varphi)^2-\frac{\varkappa-\varkappa^2\tilde{\gamma}k_0^2}{2}\dot{\varphi}^2 \\
    &-i\varkappa\nu k_0\dot{\varphi}\varphi_x - \tilde{\sigma}k_0^2\varphi_x^2\,.
    \end{split}
\end{equation}
Constant terms have been omitted from Eq.~(\ref{eqn:L_renormalized}) as they do not contribute to the temperature-dependence of the equations of state. Linear in $\varphi$ terms have been omitted because they integrate to zero over the Euclidean spacetime. Terms of order $\mathcal{O}(\varphi^3)$ and higher have been omitted because they only provide subleading corrections, and terms of order $\mathcal{O}(k_0^3)$ and higher have been omitted because $n_s$ is given only as a second derivative with respect to $k_0$.

The grand potential corresponding to the Lagrangian~(\ref{eqn:L_renormalized}) can be straightforwardly evaluated in the Fourier representation. Performing the functional integral over the quadratic field $\varphi$, for the generalized ``pressure" $\tilde{p} (\mu, T, k_0 )$ we have
\begin{equation}\label{eqn:omega}
  \tilde{p} (\mu, T, k_0 ) \, =\,   -\frac{\Omega}{V} \,=\, - \frac{T}{2V}\sum_{\mathbf{k}, \omega}\ln\epsilon(\mathbf{k},\omega)\,,
\end{equation}
where
\begin{equation}\label{eqn:epsilon_def}
    \begin{split}
    \epsilon(\mathbf{k},\omega)=\pi^{-1}[&(n_s^{(0)}+\tilde{\sigma}k_0^2)k^2+(\varkappa-\varkappa^2\tilde{\gamma}k_0^2)\omega^2 \\
    & +2i\varkappa\nu k_0\omega k_x + 2\tilde{\sigma}k_0^2k_x^2]\,.
    \end{split}
\end{equation}

\subsection{Finite-$T$ and finite-$L$ corrections}

The sum in Eq.~(\ref{eqn:omega}) is UV-divergent. Hence, we need to follow the same protocol as in the previous sections and subtract from Eq.~(\ref{eqn:omega}) the ground-state infinite-size contribution to the generalized pressure. In practice, the regularized quantity (the subscript ``$\infty$" stands for the infinite system size)
\begin{equation}
    \Delta \tilde{p}(\mu, T, k_0) = \tilde{p}(\mu, T, k_0)-\tilde{p}_\infty (\mu, 0, k_0)\,,
\end{equation}
is obtained by subtracting from Eq.~(\ref{eqn:omega}) the integral corresponding to the sum over $(\mathbf{k},\omega)$, with a matching UV-cutoff. Observe that as long as the action is quadratic in $\varphi$, the quantity $\epsilon(\mathbf{k},\omega)$ is guaranteed to be bilinear in components of $(\mathbf{k}, \omega)$. This, combined with the structure of Eq.~(\ref{eqn:omega}), implies the universal $T^{d+1}$ and 1/$L^{d+1}$ scaling of $\Delta \tilde{p}$ and, correspondingly, the same scaling of all its partial derivatives with respect to $\mu$ and $k_0$.

For instance, say we want to compute $\Delta\tilde{p}$ in an infinite system at finite temperature. Then, $\Delta\tilde{p}$ would take the form
\begin{equation}\label{eqn:Delta_p}
    \begin{split}
    \Delta\tilde{p} = -\frac{T}{2}\int\frac{d^dk}{(2\pi)^d}&\bigg[\sum_n\ln\epsilon(\mathbf{k},2\pi nT) \\
    &-\int dn\ln\epsilon(\mathbf{k},2\pi n T)\bigg]\, ,
    \end{split}
\end{equation}
where the sum and integral over $n$ have matching UV-cutoffs. Now, making the variable substitution $\mathbf{q}=\mathbf{k}/T$, the quadratic structure of $\epsilon$ allows us to pull out a factor of $\ln T^2$, which cancels exactly between the sum and the integral since they have the same UV-cutoff. We are thus left with an expression $\Delta\tilde{p}\propto T^{d+1}$, a power law inherited by thermodynamic quantities generated by $\Delta\tilde{p}$. The same argument applies when computing $\Delta\tilde{p}$ at $T=0$, with all linear system sizes infinite except for one, leading to the $1/L^{d+1}$ scaling.

The regularized generating function $\Delta\tilde{p}(\mu,T,k_0)$ now produces exactly the depletion formulas for the equations of state,
\begin{equation}
    \begin{split}
    \Delta n(\mu, T) &=\frac{\partial\Delta\tilde{p}(\mu, T, k_0=0)}{\partial\mu}, \\
    \Delta n_s(\mu, T) & = -\frac{\partial^2\Delta\tilde{p}(\mu,T,k_0)}{\partial k_0^2}\bigg|_{k_0=0}\,.
    \end{split}
\end{equation}
These can be expressed in terms of the Fourier-space representations of the correlators encountered in Sec.~\ref{sec:Quant}, yielding
\begin{equation}\label{eqn:n_correlators}
    \Delta n(\mu,T)=\frac{n_s^{(0)}\varkappa\lambda}{2}\Delta D-\frac{\nu}{2}\Delta C
\end{equation}
and
\begin{equation}\label{eqn:ns_grand_correlators}
    \Delta n_s(\mu, T) =\nu^2\Delta A+\frac{2\tilde{\sigma}}{\varkappa}\Delta B+\frac{\tilde{\sigma}}{\varkappa}\Delta C-\tilde{\gamma}n_s^{(0)}\Delta D \,,
\end{equation}
clearly satisfying Eq.~(\ref{eqn:ns_n_relation}). Although both the expressions ~(\ref{eqn:n_correlators}) and (\ref{eqn:ns_grand_correlators}) were generated by the same potential, the physics behind them are distinctly different. Since the number density is generated by the grand potential at zero superflow wavevector, it is naturally associated with the thermal fluctuations of an ideal phonon gas, and/or finite-size corrections to corresponding zero-point fluctuations. The same is not true for the superfluid stiffness. Every single term in Eq.~(\ref{eqn:ns_grand_correlators}) can be associated with a non-harmonic term in the effective action, meaning that $\Delta n_s(\mu ,T)$ has little to do with the ideal phonon gas, or more specifically, the phonon wind---as long as the system is not Galilean invariant.

\section{Numerics}
\label{sec:Num}

\subsection{Estimators}
All quantities involved in the theory can be extracted from Worm Algorithm path-integral Monte Carlo
simulations \cite{Worm,PRE} performed in the Grand canonical ensemble at low temperature in systems of finite size $L$.
[In what follows all expressions refer to systems with the same size in all dimensions; i.e. the system volume is $V=L^d$.]
If $P(N,\mathbf{M})$ is the probability of having a state with  the
particle number $N$, and winding number $\mathbf{M}$ (it is a $d$-dimensional vector) then
\be
n=V^{-1} \sum_{N \mathbf{M}} N P(N,\mathbf{M}) \equiv \frac{1}{V}\, \langle N\rangle \,,
\label{Kdensity}
\ee
\be
\varkappa= \frac{1}{TV} \, \langle [N- \langle N\rangle ]^2\rangle \,,
\label{Kcompressibility}
\ee
\be
n_s = \frac{TL^{2-d}}{d}\, \langle \mathbf{M}^2\rangle \,.
\label{Kstiffness}
\ee

While it is possible to derive estimators also for $\nu$ and $\gamma$, in practice it is much easier to estimate them in a large system by performing
a numerical derivative based on simulations at different densities.
Numerical derivative is obviously less accurate than $n_s$ itself, but there is no problem in
obtaining it with a few percent accuracy.

Turning to the estimator for the parameter $\sigma$, we note that the estimator
\be
\tilde {\sigma} = \frac{TL^{4-d}}{6d} \, \sum_{\alpha =1}^{d} \, 
 \left[ \,  3 \langle \mathbf{M}_{\alpha}^2\rangle^2  -  \langle \mathbf{M}_{\alpha}^4\rangle  \, \right] 
\label{sigma_wind}
\ee
[a direct analog of the estimator (\ref{sigma_W}) for the parameter $\theta$ of the classical model] produces a somewhat different quantity. This is because the worm algorithm used for simulations works in the grand canonical ensemble. In this case, we get 
\be
\tilde{\sigma} \, =\, {2\over V}  {\partial^2 \Omega (\mu, k_0^2)\over \partial(k_0^2)^2}\bigg|_{k_0=0} \, =\,  {\partial n_s(\mu, k_0^2)\over \partial (k_0^2)}\bigg|_{k_0=0} 
\ee
($\Omega$ is the grand canonical potential, $k_0$ has the same meaning as before), which is different from
\be
\sigma \, =\, {\partial n_s (n, k_0^2)\over \partial (k_0^2)}\bigg|_{k_0=0}\,.
\ee
As shown in Appendix~\ref{appendix:action}, they are related to each other through
\begin{equation}
    \tilde{\sigma}=\sigma-\varkappa\nu^2/2\,.
\end{equation}

The estimator for the parameter $\tilde{\sigma}$ also involves precise cancellation  of large terms.
Given that winding number fluctuations themselves scale according to (\ref{Kstiffness}),
the required relative accuracy of computing each average in (\ref{sigma_wind}) must be much smaller than
\be
 \epsilon \sim \frac{\tilde{\sigma}} {TL^{4-d}}\left(  \frac{mTL^{2-d}}{n_s} \right)^2
 \sim \frac{\tilde{\sigma} m^2}{n_s^2} \, \frac{1}{\beta V}\,.
\label{sigma_eps}
\ee
This result implies that even if simulations are based on a ``perfect" Monte Carlo algorithm
(with autocorrelation time equal to one sweep, or $ \propto V \beta $ updates)
the required number of updates scales as the $(d+1)$-dimensional system volume cubed. Fortunately,
in this work we are interested in properties of the superfluid phase close to the ground state
and thus small system sizes $L<10$ can be used for computing $\tilde{\sigma}$.

\subsection{Model}
In our simulations, we employ the microscopic Hamiltonian
\begin{equation}\label{eqn:bose_hubbard}
    H = -t\sum_{\langle i,j\rangle}(c_i^{\dagger}c_j+\text{H.c.})+U\sum_in_i^2-\mu\sum_in_i \, ,
\end{equation}
of the Bose-Hubbard model. Here   $c_i^{\dagger}$ and $c_i$ are bosonic creation and annihilation operators respectively, and $n_i=c_i^{\dagger}c_i$. All simulations are performed in two spatial dimensions on the square lattice. At low temperature the Hamiltonian~(\ref{eqn:bose_hubbard}) exhibits superfluidity. Because of the lack of continuous translation invariance, the depletion will not be described by Landau theory, but instead by the more general expression~(\ref{Delta_n_s_gen}).

\subsection{Soft-core bosons}
\begin{figure}
    \centering
    \includegraphics[width=0.9\linewidth]{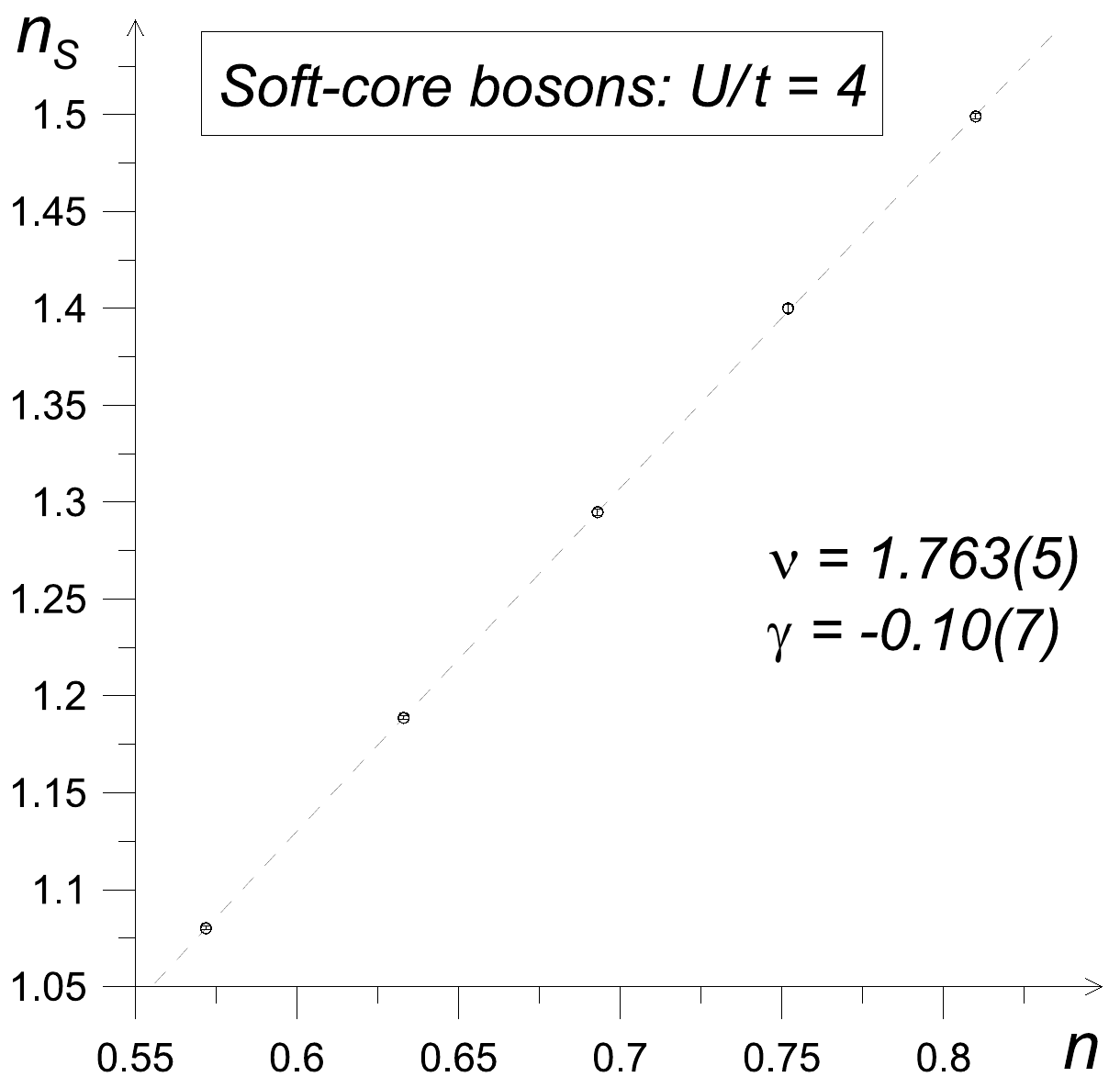}
    \caption{ Superfluid stiffness as a function of density for soft-core bosons
    in 2D with $U/t=4$ (for system size $L\times L=64\times64$ and $\beta / t =6$). 
    Dashed curve is a parabolic fit with $\nu=1.763(5)ta^2$ and $\gamma=0.10(7)ta^4$. 
    Error bars are smaller than symbol sizes.
    }
    \label{fig:nsnsoft}
\end{figure}

At generic filling factor and at moderate values of on-site repulsion, we have a system 
of superfluid soft-core bosons. In this case, the largest parameter controlling the superfluid density depletion with temperature in Eq.~(\ref{eqn:delta_ns_final}) 
is coming from large $\nu$. From the accurate data for $n_s(n)$ dependence, see Fig.~\ref{fig:nsnsoft},  we extract the value of $\nu$, which in this case is 
large but still significantly different from the Galilean invariant result $\nu=2ta^2$.
In sharp contrast with the hard-core system considered in the next subsection, 
the coefficient $\gamma$ happens to be very small for this parameter 
regime and within the accuracy of our simulations may be set to zero.  

\begin{figure}
    \centering
    \includegraphics[width=0.9\linewidth]{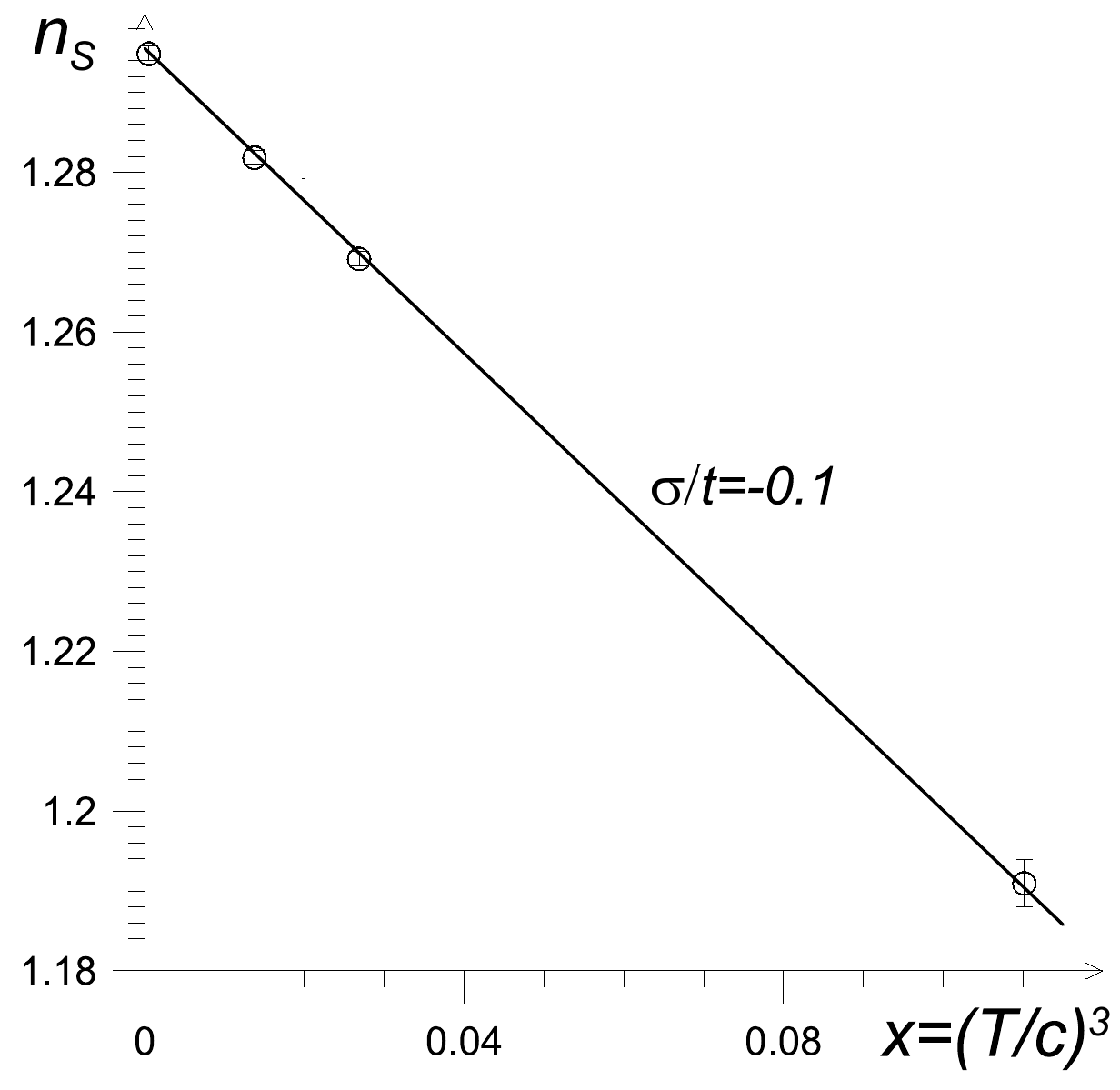}
    \caption{Superfluid stiffness depletion as a function of $x=(cT)^3$ 
    for soft-core bosons in 2D with $U/t=4$ (for system size $L\times L=64\times64$).
    Solid (black) curve is the $f(x) = A + Bx$ theoretical prediction 
    with $A=n_s(T=t/6)$ and $B$ fixed by the measured system parameters according to Eq.~(\ref{eqn:delta_ns_final}). 
    }
    \label{fig:nsT3soft}
\end{figure}

For study of finite-temperature effects, we consider system density $n \approx 0.693$
characterized by compressibility $\varkappa = 0.2974(5) t$ and sound velocity 
$c=2.087(3) ta$.
It turned out that simulations of soft-core bosons are very demanding: the 
$\tilde{\sigma}$ parameter can be computed only for system sizes $L\leq 6$. The estimate of $\tilde{\sigma}$ (using small system sizes)
from moments of the winding number distribution, Eq.~(\ref{sigma_wind}), is 
$\tilde{\sigma} (L=6) \approx -0.60(10)t$. 
This result implies that $\sigma = -0.15(10)$.
In Fig.~\ref{fig:nsT3soft} we compare computed $n_s(T)$
dependence with the fit-free theoretical prediction based on Eq.~(\ref{eqn:delta_ns_final}) and $\sigma =-0.1$.
The agreement is near perfect, i.e. within error bars for all points.

\subsection{Hard-core bosons}

\begin{figure}
    \centering
    \includegraphics[width=0.9\linewidth]{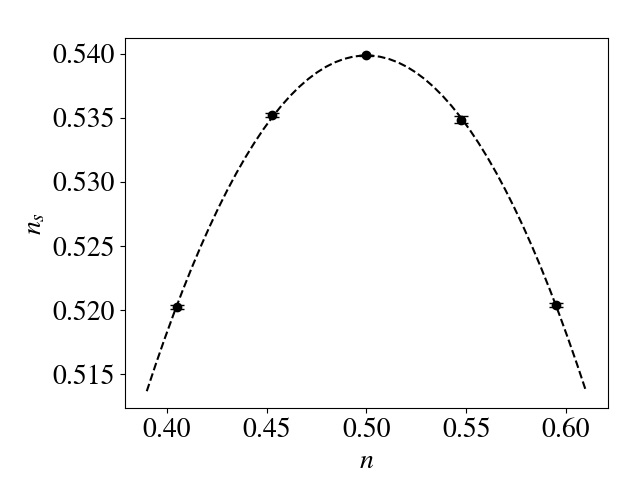}
    \caption{Superfluid stiffness as a function of density for a hard-core boson system with $L=c\beta=10$. Parameter $\gamma$ is calculated with a quadratic fit, which yields $\gamma=-2.14(1)ta^4$ when extrapolated to $T=0$.}
    \label{fig:hc_gamma}
\end{figure}

If the limit $U\rightarrow\infty$ is taken in the Bose-Hubbard Hamiltonian~(\ref{eqn:bose_hubbard}), we acquire the so-called \textit{hard-core boson} model. At half-filling ($\mu=0$), this model is particle-hole symmetric, meaning that the parameter $\nu$ appearing in Eq.~(\ref{eqn:delta_ns_final}) is identically zero. Unlike the soft-core bosons, the parameter $\gamma$ is not negligible in the hard-core system. We extract $\gamma$ by fitting a quadratic to $n_s(n)$ data as shown in Fig.~\ref{fig:hc_gamma}, yielding $\gamma=-2.14(1)ta^4$ when extrapolating to $T=0$. Using the estimators in Sec.~\ref{sec:Num}, we compute for the hard-core system $c=2.2688(3)ta$ and $\varkappa=0.10476(1)t$. 

For the hard-core system, deviations from the $T^3$-law happen at lower temperatures than in the soft-core system, motivating a more detailed analysis of the finite-temperature depletion of $n_s$. As with the classical system, we will proceed by fitting our data not only with the theoretical $T^3$-prediction, but also with a subleading correction proportional to $T^5$. Simulating a reliable value of the parameter $\sigma$ proves once again to be a non-trivial task. Here, instead of trying to extrapolate $\sigma$ from simulations of small systems, we adopt a different approach: Since depletion of $n_s$ in an isotropic system depends on the two parameters $\beta$ and $\lambda_{\beta}=c\beta/L$ (see Sec.~\ref{sec:fin_T_L}), we choose to simulate two different systems, with different values of $\lambda_{\beta}$. That is to say, while varying temperature in the simulations, we also vary system size such that $\lambda_{\beta}$ is fixed to the same value for all data points. By manually varying $\sigma$ and the ground-state value $n_s(T=0)$ (which should be the same for both systems), we can verify the theory by finding values of $\sigma$ and $n_s(T=0)$ that can predict the depletion in both cases.

\begin{figure}
    \centering
    \includegraphics[width=0.9\linewidth]{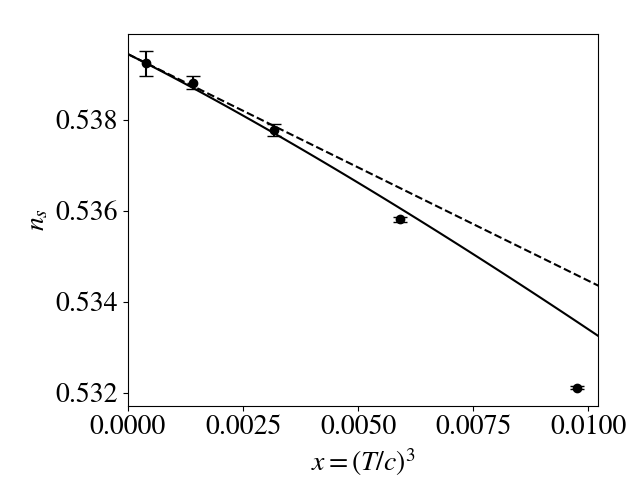}
    \caption{Finite-temperature depletion of superfluid stiffness in the 2D system of hard-core bosons. System size was chosen such that the aspect ratio $L/c\beta\approx4.698$ for all data points. Solid line is a fit of the function $f(x)=A_0+A_1x+A_2x^{5/3}$, where $A_0=0.53944$, $A_2=-2.3$ are fitting parameters, and $A_1$ is held fixed by the measured system parameters with $\sigma=-0.107$. Dashed line is the same function but with $A_2$ set to zero, showing the asymptotic behavior at the approach to $T=0$. The correlator $c(c\beta)^3\Delta B\approx~0.1686$ was computed numerically using Eq.~(\ref{eqn:B_nondim}) for the given aspect ratio, with $\Delta C$ and $\Delta D$ then given by Eq.~(\ref{eqn:C_nondim}) and Eq.~(\ref{eqn:D_nondim}) respectively.}
    \label{fig:T_depletion}
\end{figure}

\begin{figure}
    \centering
    \includegraphics[width=0.9\linewidth]{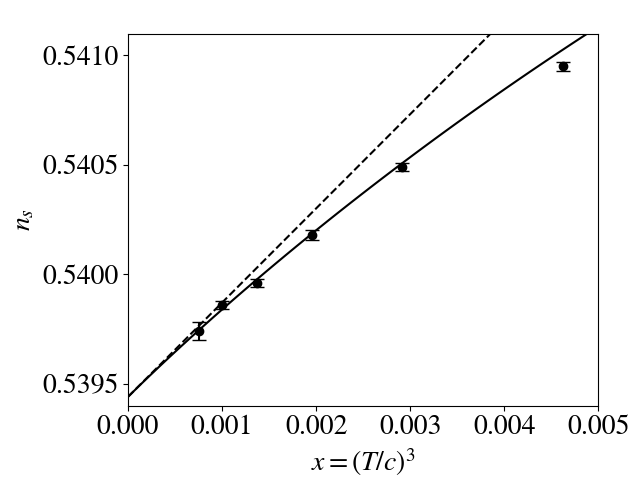}
    \caption{The interplay of finite-$\beta$ and finite-size effects on the superfluid stiffness in a square system of hard-core bosons at half-filling. System size is such that $c\beta=L$. As suggested by Eq.~(\ref{st_hypercube}), the stiffness increases as the temperature increases because of the simultaneous decrease of the system size. Solid line is a fit of the form $A_0 + A_1x + A_2x^{5/3}$, where $x=(T/c)^3$, and $A_1$ is fixed by Eq.~(\ref{st_hypercube}) with $\sigma=-0.107$. $A_0$ and $A_2$ are fitting parameters, with the fit yielding $A_0=0.53944$ and $A_2=-3.15$. Dashed line is the same curve with $A_2$ set to zero, showing the low-temperature $T^3$ law predicted by the theory.}
    \label{fig:st_hypercube}
\end{figure}

\begin{figure}
    \centering
    \includegraphics[width=0.9\linewidth]{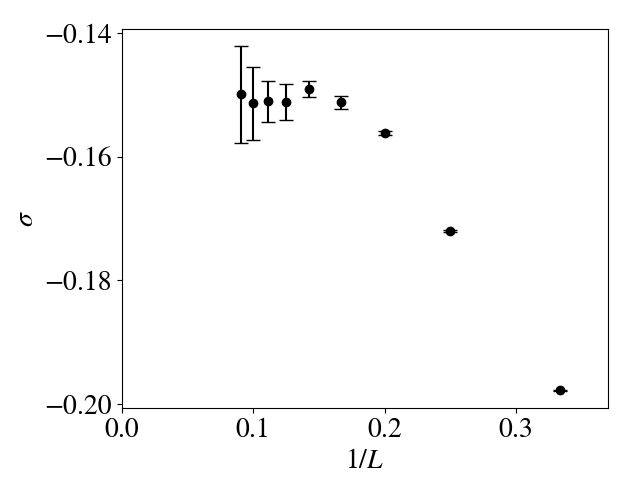}
    \caption{Winding number estimator~(\ref{sigma_wind}) for $\sigma$ as a function of inverse system size. Temperature is chosen such that $c\beta=L$ for each data point. There is an apparent plateau at $\sigma\approx-0.15$. However, the strong finite-size effects and large statistical uncertainties prevent a reliable $L\rightarrow\infty$ extrapolation.}
    \label{fig:sigma_vs_L}
\end{figure}

Figure~\ref{fig:T_depletion} shows $n_s$ as a function of $(T/c)^3$ for $L\approx4.698c\beta$, and Fig.~\ref{fig:st_hypercube} shows the same for $L=c\beta$. While the data in Fig.~\ref{fig:T_depletion} could in principle be well described by the asymptotic formula (\ref{Delta_n_s_fin}), in the interest of maximal accuracy, we opt instead to compute $\Delta n_s$ using the results of Sec.~\ref{sec:fin_T_L}. For $L\approx4.698c\beta$ we find numerically $c(c\beta)^3\Delta B\approx0.1686$, with Eqs.~(\ref{eqn:C_nondim}) and (\ref{eqn:D_nondim}) yielding the remaining correlators. For the $L=c\beta$ case, the exact analytical result is given by Eq.~(\ref{st_hypercube}). We find that both sets of data are fitted well when $\sigma=-0.107$. This is in slight contrast to the available data from the winding number estimator~(\ref{sigma_wind}), shown in Fig.~\ref{fig:sigma_vs_L}, which seems to indicate $\sigma\approx-0.15$. However, the significant statistical errors in the winding number data makes it hard to extrapolate a reliable value for $\sigma$, as it is not certain if the trend changes at larger system sizes or not. In principle it is also possible to extract $\sigma$ from Eq.~(\ref{eqn:sigma_q_anis}), similarly to what was done for the classical system. It turns out not to be as practical for the quantum system however, due to significant finite-size effects preventing any reliable extrapolation from the few accessible data points.

\section{Conclusion}
\label{sec:disc}

We developed a field-theoretical framework for calculating finite-size corrections to harmonic terms of a U(1)-symmetric effective action. The corrections are due to truncation of the renormalization of the harmonic terms by the non-harmonic ones. We applied our theory to the problem of the low-temperature depletion of superfluid stiffness in a generic superfluid.
The depletion is controlled by the three parameters: $\nu$, $\gamma$, and $\sigma$, with the result expressed by Eqs.~(\ref{eqn:delta_ns_final})--(\ref{eqn:nu}). In a Galilean system, parameters $\gamma$  and $\sigma$ are identically equal to zero and $\nu$ equals to the inverse particle mass.  This is how Eqs.~(\ref{eqn:delta_ns_final})--(\ref{eqn:nu}) recover Landau's relation, while simultaneously revealing its deficiency in a generic case. 

At the qualitative level, our analysis predicts phenomena fundamentally absent from the Landau theory. While a Galilean system will have a ground-state superfluid density exactly equal to the total density at all system sizes, in the general case, one will see finite-size corrections described by Eqs.~(\ref{Delta_n_s_long}) and (\ref{Delta_n_s_trans}). The form of Eq.~(\ref{eqn:delta_ns_final}) also suggests that there might exist exotic regimes where one can see {\it enhancement} of the superfluid density instead of depletion. 

It is important to realize that even at $\gamma = \sigma = 0$, the requirement $\nu = 1/m$ is {\it necessary} for the depletion of the superfluid density $-\Delta \rho_s=-m \Delta n_s$ to coincide with the ``normal density" of the phonon wind, $n_{\rm wind}$. Based on the definition 
(with $v_0$ the velocity of the wind)
\be
n_{\rm wind} = \lim_{v_0 \to 0} |\langle {\bf j} \rangle|/v_0 \, ,
\label{n_wind}
\ee
we readily calculate $n_{\rm wind}$ from the harmonic part of the hydrodynamic Hamiltonian. The leading contribution comes from the second term in the r.h.s. of  (\ref{eqn:j_hydro}) yielding the relation between $-\Delta \rho_s$ and $n_{\rm wind}$:
\be
- \Delta \rho_s \, =\, \nu m  n_{\rm wind} \qquad (\gamma = \sigma = 0) \, .
\label{comparison}
\ee
This way we prove that in a general case, the phonon wind and thermal depletion of superfluid density are fundamentally different phenomena.

The fact that our theory reproduces the result of Landau should be considered as an important and nontrivial check point. Further and more general validation is provided by numeric simulations. The agreement with numeric results is satisfactory but there is still room for further work. The main challenge is associated with direct estimation of parameter $\sigma$. Small system sizes are not sufficient because of large systematic error, while for linear system sizes larger than 10 the signal becomes smaller than statistical noise. It might also be useful to further develop theory to understand the scaling of subleading corrections to the finite-temperature (finite-size) effects to be able to fit numeric data---clearly showing the importance of subleading terms---with corresponding ansatz.

The broader context of our approach is the universal low-temperature thermodynamics of superfluids characterized by an intrinsic connection between finite-$T$ and finite-$L$ effects and implying universal scaling, $T^{d+1}$ and $1/L^{d+1}$, respectively, for a large class of thermodynamic quantities, including superfluid density, in both canonical and grand canonical ensembles. This prediction has been compared to existing experimental literature in a companion letter \cite{companion}.

\begin{acknowledgments}
This work was stimulated by numerous conversations (though without reaching a consensus \cite{hegg2024universallowtemperaturefluctuationunconventional}) with Wei Ku about the scientific status of Landau theory of low-temperature depletion of superfluid density. 
This work was supported by the National Science Foundation under Grant DMR-2335904.
\end{acknowledgments}

\bibliography{refs.bib}

\clearpage
\appendix

\section{Useful mathematical relations}\label{subsec:math}
Each expression for regularized correlators of interest 
can be reduced to the evaluation of some dimensionless constant, $\alpha$, given by a $d$-dimensional integral over ${\bf q}$ with 
an integrand being a regularized sum, 
$\Sigma (q)$, times $q$ raised to a certain integer power $l$:
\be
\alpha \, =\, \int {{\rm d} ^d q \over (2\pi)^d}\,  q^l \Sigma (q)\, ,
\label{alpha}
\ee
\be
\Sigma (q) \, =\, \lim_{n_*\to \infty} \left[ \sum_{n=-n_*}^{n_*} f(q,n)  - \int_{n=-n_*}^{n_*}  \!\!\!  f(q,n)  dn \right]\, .
\label{S_q}
\ee

As described above, the regularization of the sum amounts to subtracting the corresponding integral.
If the sum and integral converge, then $n_*$ can be replaced with $\infty$ in the sum and integral individually.

The first case of this type is 
\be
\alpha_1 \, =\, \int {{\rm d} ^d q \over (2\pi)^d}\,  q^2 \Sigma_1(q)\, ,
\label{alpha_1_def}
\ee
\be
f_1(q,n) \, =\, {1\over (2\pi n)^2 + q^2} \, .
\label{f_1}
\ee
Using standard mathematical result
\be
\sum_{n=-\infty}^{\infty}  {1\over (2\pi n^2) + q^2} \, =\, {1\over 2q } \coth(q/2) \, ,
\label{standard}
\ee
we arrive at
\be
\alpha_1 \, =\, \int {{\rm d} ^d q \over (2\pi)^d}\, {q\over e^q -1} \, = \, I_d \, .
\label{alpha_1_res}
\ee

The second case is
\be
\alpha_2 \, =\, \int {{\rm d} ^d q \over (2\pi)^d}\,  q^4 \Sigma_2(q)\, ,
\label{alpha_2_def}
\ee
\be
f_2(q,n) \, =\, {1 \over [(2\pi n)^2 + q^2]^2 } \, .
\label{f_1}
\ee
Here we observed that
\[
\Sigma_2(q) \, =\, -{1\over 2q} {{\rm d} \Sigma_1(q) \over {\rm d}  q} \, ,
\]
which allows us to relate $\alpha_2$ to $\alpha_1$ by doing the integral in (\ref{alpha_2_def}) by parts:
\be
\alpha_2 \, =\, {d+2 \over 2} \, \alpha_1 \, =\,  {d+2 \over 2} \, I_d \, .
\label{alpha_2_res}
\ee

At the first glance, the third case    
\be
\alpha_3 \, =\, \int {{\rm d} ^d q \over (2\pi)^d}\,  q^2 \Sigma_3(q)\, ,
\label{alpha_3_def}
\ee

\be
f_3(q,n) \, =\, {(2\pi n)^2 \over [(2\pi n)^2 + q^2]^2 } \, ,
\label{f_3}
\ee
requires careful regularization. However, by observing that
\[
f_3(q,n) \, =\, {1 \over (2\pi n)^2 + q^2  } \, -\, {q ^2 \over [(2\pi n)^2 + q^2]^2 } \, ,
\]
we immediately conclude that
\be
\alpha_3 \, =\,  \alpha_1 - \alpha_2 \, =\,  - {d\over 2}  \, I_d \,.
\label{alpha_3_res}
\ee

The fourth case 
\be
\alpha_4 \, =\, \int {{\rm d} ^d q \over (2\pi)^d}\,   \Sigma_4 (q)\, ,
\label{alpha_4_def}
\ee
\be
f_4(q,n) \, =\, {(2\pi n)^2 \over (2\pi n)^2 + q^2 } \, ,
\label{f_4}
\ee
is processed similarly. Here we observe that
\[
f_4(q,n) \, =\, 1 \, -\, {q ^2 \over (2\pi n)^2 + q^2  } \, .
\]
The first term in the r.h.s. is $n$-independent and thus can be  omitted because of exact cancellation between the sum and the regularizing integral.
We conclude that
\be
\alpha_4 \, =\, -\alpha_1 \, =\, -I_d.
\label{alpha_4_res}
\ee

\section{Symmetrization of integrals}\label{appendix_symm}
The substitution $x_i^2\rightarrow r^2/d$ in an integral, where $r^2=\sum_ix_i^2$, is trivial given that the expression multiplying $x_i^2$ in the integrand is an isotropic function. This relation can be generalized to higher order (in $x_i)$ polynomials as follows:

Let ${\bf r}=(x_1, \ldots, x_d)$ be a $d$-dimensional vector and $f(r)$ some function that depends only on the magnitude $r$ of the vector ${\bf r}$.  Consider an integral
\be
I \, =\, \int d^d r \left( \prod_{j=1}^d x_j^{2\alpha_j} \right) f(r) \, ,
\label{I}
\ee
where $\alpha_j$'s are certain nonnegative integers. Integrating first over the surface of the hypersphere of the hyperradius $r$, 
\[
\int_{{\cal S}_r} dS \prod_{j=1}^d x_j^{2\alpha_j} 
\]
and observing that [below $S(r)$ is the hyper-area of the hypersphere ${\cal S}_r$]
\[
\Theta \, =\, {1\over S(r) \, r^{2\alpha_*}} \int_{{\cal S}_r} dS \prod_{j=1}^d x_j^{2\alpha_j}  \qquad 
\left( \alpha_* \, =\, \sum_{j=1}^d \alpha_j \right)
\]
is an $r$-independent function of $\vec{\alpha} = (\alpha_1, \ldots, \alpha_d)$, we conclude that the integral (\ref{I}) reduces to the following isotropic form
\be
I \, =\, \Theta (\vec{\alpha})  \int  r^{2\alpha_*} f(r) \, d^d r \, .
\label{II}
\ee
The function $\Theta (\vec{\alpha})$ is readily found by explicitly calculating the integral $I_0$ corresponding to $f(r) = e^{-r^2}$. On the one hand, using the original form (\ref{I}), we have
\[
I_0  = \! (-1)^{\alpha_*} \!\! \left(\!\!  \prod_{\scriptsize \begin{array}{c} j=1 \\
\alpha_j \neq 0 \end{array} }^d \!\!\! {\partial^{\alpha_j} \over \partial \lambda_j^{\alpha_j}}
\! \right) \! \left.  \int\!  d^d r \exp \left( \! - \sum_{j=1}^d \lambda_j x_j^2 \right) \right|_{\lambda_j =1} \! \!  .
\]
On the other hand, from the representation (\ref{II}) we are supposed to have
\[
I_0  = \Theta (\vec{\alpha}) \left. (-1)^{\alpha_*}  {{\rm d}^{\alpha_*} \over {\rm d}\lambda^{\alpha_*}}  \! \int\!  d^d r \exp \left(- \lambda \sum_{j=1}^d  x_j^2 \right) \right|_{\lambda  =1} .
\]
Now explicitly performing the Gaussian integrals in both expressions and then requiring that the two results for $I_0$ be equal, we find
\be
\Theta (\vec{\alpha}) = \left( \left. {{\rm d}^{\alpha_*} \lambda^{-d/2} \over {\rm d} \lambda^{\alpha_*}} \right|_{\lambda=1} \right)^{\!\! -1}  \!\!\!\!  \prod_{\scriptsize \begin{array}{c} j=1 \\
\alpha_j \neq 0 \end{array} }^d \!\!\! \left( \left.
{{\rm d}^{\alpha_j} \lambda^{-1/2} \over {\rm d}\lambda^{\alpha_j}}\right|_{\lambda =1} \right)
.
\ee

\section{Expansion coefficients in phase-only representation}\label{appendix:action}
All expansion coefficients in the effective Lagrangian can be written as derivatives of the generalized ground-state pressure $p_0(\mu, k_0^2)$---the negative density of the zero-point grand canonical potential---with respect to either chemical potential or square of superflow wavevector. However, one can also write them in terms of the expansion coefficients appearing in the hydrodynamic Hamiltonian. Writing the effective Lagrangian generally as
\begin{equation}
    \begin{split}
        \mathcal{L}(\Phi)=&\xi_1(\nabla\Phi)^2-\xi_2\dot{\Phi}^2-i\xi_3\dot{\Phi}(\nabla\Phi)^2 \\
        &-\xi_4\dot{\Phi}^2(\nabla\Phi)^2 + \xi_5(\nabla\Phi)^4\,,
    \end{split}
\end{equation}
we proceed by determining the coefficients $\xi_i$ one by one. The coefficients of the harmonic terms are standard quantities found for instance in Ref.~\cite{book}. We have
\begin{equation}
    \xi_1=\frac{\partial p_0(\mu,k_0^2)}{\partial(k_0^2)}=-\frac{n_s}{2}
\end{equation}
and
\begin{equation}
    \xi_2=\frac{1}{2}\frac{\partial^2p_0(\mu,k_0^2)}{\partial\mu^2}=\frac{\varkappa}{2}\,.
\end{equation}
The next coefficient $\xi_3$ is then given by
\begin{equation}
    \xi_3=\frac{\partial\xi_1}{\partial\mu}=\frac{\varkappa}{2}\frac{\partial n_s}{\partial n}=\frac{\varkappa\nu}{2}.
\end{equation}
Coefficient $\xi_4$ is given by the derivative of $\xi_3$. Using Eq.~(\ref{eqn:lambda_derivative}), we have
\begin{equation}
    \begin{split}
    \xi_4=\frac{1}{2}\frac{\partial\xi_3}{\partial\mu}&=\frac{1}{4}\bigg(\frac{\partial\varkappa}{\partial\mu}\nu+\varkappa^2\frac{\partial\nu}{\partial n}\bigg) \\
    &=\frac{\varkappa^2}{2}\bigg(\gamma-\frac{\varkappa\nu\lambda}{2}\bigg)\,.
    \end{split}
\end{equation}
Determining the final coefficient $\xi_5$ is most conveniently done with the Jacobian technique:
\[
{\partial n_s (\mu, k_0^2)\over \partial (k_0^2)}  =  {{\cal D} (n_s,\mu)\over {\cal D}(k_0^2,\mu)}  = {{\cal D} (n_s,\mu)\over {\cal D}(k_0^2, n)} \left[ { {\cal D}(k_0^2,\mu)} \over {\cal D}(k_0^2, n)\right]^{-1} \!\! .
\]
Observing/recalling that
\[
{\partial n_s(n, k_0^2) \over \partial n}\bigg|_{k_0=0}= \, \nu \, ,  \qquad {\partial n_s(n, k_0^2) \over \partial (k_0^2)}\bigg|_{k_0=0}= \, \sigma \, , 
\]
\bea
{\partial \mu (n, k_0^2) \over \partial (k_0^2)}\bigg|_{k_0=0} &=& {\partial^2 {\cal E} (n,k_0^2) \over \partial n \,  \partial (k_0^2)} \bigg|_{k_0=0} \nonumber \\
&=& {1\over 2}{\partial n_s (n,k_0^2) \over  \,  \partial n} \bigg|_{k_0=0}  \, =\, \, {\nu\over 2} \, , \nonumber
\eea
\[
{\partial \mu (n, k_0^2) \over \partial n}\bigg|_{k_0=0} = \, \varkappa^{-1}\, , 
\]
\[
\left[ { {\cal D}(k_0^2,\mu)} \over {\cal D}(k_0^2, n)\right]^{-1} \, =\, {\partial n(\mu, k_0^2)\over \partial \mu} \bigg|_{k_0=0}= \, \varkappa \, ,
\]
we conclude that 
\be
\xi_5 \, =\, -\frac{1}{4}(\sigma - \varkappa \nu^2/2) \, .
\ee
\end{document}